\documentclass[12pt,preprint]{aastex}

\shorttitle{Clumpy H$_2$ Shell of R Aqr}
\shortauthors{Ragland et al.}

\begin{document}


\title{First Images of R Aquarii and its Asymmetric H$_{2}$O Shell}


\author{Ragland, S.\altaffilmark{1}, 
Le Coroller, H.\altaffilmark{1,2},
Pluzhnik, E. \altaffilmark{1,10},
Cotton, W. D.\altaffilmark{3}, 
Danchi, W.C.\altaffilmark{4},
Monnier, J. D.\altaffilmark{5},
Traub, W.A.\altaffilmark{6},
Willson, L. A.\altaffilmark{7},
Berger, J.-P.\altaffilmark{8},
Lacasse, M. G.\altaffilmark{9},
}

\altaffiltext{1}{California Association for Research in Astronomy, 65-1120 Mamalahoa Hwy, Kamuela, HI 96743; sragland@keck.hawaii.edu}
\altaffiltext{2}{Currently at Observatoire de
Haute-Provence, OHP/CNRS, F-04870 St.Michel l'Observatoire, France}
\altaffiltext{3}{National Radio Astronomy Observatory, 520 Edgemont Road, 
Charlottesville, VA 22903}
\altaffiltext{4}{NASA Goddard Space Flight Center, Exoplanets \& Stellar 
Astrophysics, Code 667, Greenbelt, MD 20771}
\altaffiltext{5}{University of Michigan at Ann Arbor, Department of 
Astronomy, 500 Church Street, Ann Arbor, MI 48109-1090.}
\altaffiltext{6}{Jet Propulsion Laboratory,  M/S 301-451, 4800 Oak Grove Dr., Pasadena CA, 91109}
\altaffiltext{7}{Department of Physics and Astronomy, Iowa State University,
Ames IA  50014}
\altaffiltext{8}{Laboratoire d'Astrophysique de Grenoble, 414 Rue de la 
Piscine, F-38400 Saint Martin d'Heres, France.}
\altaffiltext{9}{Harvard-Smithsonian Center for Astrophysics, 60 Garden 
Street, Cambridge, MA 02138}
\altaffiltext{10}{Institute of Astronomy of Kharkov University, Ukraine.}


\begin{abstract}

We report imaging observations of the symbotic long-period Mira variable R Aquarii (R Aqr) at near-infrared and radio wavelengths. The near-infrared observations were made with the IOTA imaging interferometer in three narrow-band filters centered at 1.51, 1.64, and 1.78 $\mu$m, which sample mainly water, continuum, and water features, respectively.  Our near-infrared fringe visibility and closure phase data are analyzed using three models. (a) A uniform disk model with wavelength-dependent sizes fails to fit the visibility data, and is inconsistent with the closure phase data. (b) A three-component model, comprising a Mira star, water shell, and an off-axis point source, provide a good fit to all data.  (c)  A model generated by a constrained image reconstruction analysis provides more insight, suggesting that the water shell is highly non-uniform, i.e., clumpy.  The VLBA observations of SiO masers in the outer molecular envelope show evidence of turbulence, with jet-like features containing velocity gradients.  
\end{abstract}




\section{Introduction}

R Aqr is the brightest known symbiotic system \citep{Whitelock83} 
consisting of a Mira primary and a white dwarf companion. 
The companion has not been directly detected yet because of the 
presence of nebular emissions and circumstellar material around R Aqr 
itself. The pulsation period of 
R Aqr is 387 days. The light curves show reduced amplitude lasting 
for several cycles \citep{Mattei79}. 

Miras with close companion white dwarfs usually are classified as 
symbiotic systems \citep{Allen84, Whitelock87, Luthardt92, Bel00}.   
A few Miras are known to have companions but are not 
(or are only very mildly) symbiotic systems; this 
includes {\it o} Cet = Mira, with a probable WD 
companion in a multi-century orbit \citep{Reimers85, Wood04}. 

The occasional appearance of bright 
emission lines in the visible spectrum signifies the 
presence 
of a hot companion around R Aqr \citep{Merrill21,Merrill50}. 
The observed wide and flat minima of the light curve of R Aqr 
was interpreted as the signature of 
a companion around this target \citep{Merrill56}.
Ultraviolet observations from IUE have been interpreted as evidence of a 
white dwarf companion \citep{Michal80}.
The observed broad infrared excess at 12$\mu$m \citep{Kenyon88} could also be 
interpreted as evidence for the presence of dust heated by a hot
companion; alternatively,  
dust shells with different particle sizes or composition could also explain 
the observed broad infrared excess at 12$\mu$m. 

An early suggestion for an orbital period of 27 yrs from 
radial velocity measurements \citep{Merrill50} was not supported by later 
observations. The observed decrease in the brightness of the visible and near-infrared 
light curves of R Aqr during 1975-78 and on two
previous occasions was interpreted in terms of an obscuring 
circumstellar dust clump eclipsing the target \citep{Willson81}. These authors suggested that R Aqr is an eclipsing binary system with an orbital period 
of 44 years. Subsequently, \citet{Whitelock83} found a similar eclipsing signature on 6 symbiotic stars including R Aqr, however, 
the frequency of occurrence of these decreases in symbiotic stars (3 or 4 out of 6
well studied cases) with such long periods suggests that eclipsing by a secondary body probably is not the likely 
mechanism causing the observed long term amplitude changes of a few tens of 
years \citep{Feast83}. However, if the Mira wind geometry is altered by the
presence of the white dwarf, or by two winds colliding, then the "eclipse" could correspond
to observing through the unmodified Mira wind on the side away from the white dwarf. Alternatively, episodic dust emission observed around Mira and AGB stars \citep{Danchi95, Bester96} could explain these observed brightness modulations. A nova explosion on the surface of the white dwarf companion 
followed by dust condensation could also explain the observed long term amplitude 
changes. R Aqr could very well be a potential candidate of becoming a slow
nova such as RR Tel which is a symbiotic system with a Mira primary and infrared excess. 

In this paper we use the word ``asymmetry'' to mean that part of the 
2-dimensional brightness distribution which cannot be made symmetric with 
respect to a reflection through a point. Thus, for example, an elliptical 
uniform disc or an equal-brightness binary system are both symmetric, but 
a binary with unequal brightness or a star with an off-centered bright/dark spot 
is asymmetric \citep{Ragland06}.

Departures from circular symmetry are known in AGB stars from various 
high angular resolution observations \citep{Karovska91, Wilson92, Haniff92, Richichi95, Ragland96, Weigelt96, Karovska97, Tuthill97, Lattanzi97, Wittkowski98, Tuthill99, Tuthill00, Hofmann00, Thompson02, Monnier04, Weiner06, Ragland06}. However, no consensus exists as to the 
mechanism that would cause such departures from apparent circular symmetry 
\citep{Ragland06}. 

Observations of dust shells around AGB stars in the infrared 
\citep{Danchi94, Lopez97, Tuthill00, Monnier00, Weiner06} 
and SiO maser shells at radio/millimeter wavelengths
\citep{Diamond94, Green95, Diamond03, Cotton04, Soria04, Cotton06} also show asymmetries. 
Again, the connection between apparent 
surface features (or companions) and the morphology of the dust or SiO shells 
has not been established. 

Recent studies of astrophysical jets around a few
AGB stars \citep{Kellogg01, Imai02, Sahai03, Sokoloski03, Brocksopp04} 
from radio, X-ray or Hubble Space Telescope (HST) observations suggest 
that those stars showing substantial asymmetry may all have a low mass 
stellar companion accreting mass from the AGB primary.

The R Aqr system has complex nebular structures at arcsecond and arcminute scales. 
In addition, X-ray, UV, visible and radio observations show astrophysical jets \citep{Kellogg01, Hollis91, Solf85, Hollis85}. These large scale structures 
have no influence on our high angular resolution multi-wavelength observations 
of the central region (size $\sim 0.4$ arcsec) of this complex
system reported in this paper. 

Recently, we reported the results from the first phase of our program in which we 
surveyed 56 evolved giants including 35 Mira stars, 18 SR variables and 
3 Irr variables looking for asymmetry in their brightness profiles (\citep{Ragland06}, hereafter Paper I). 
We found that about 29\% of the AGB stars show non-zero (or non-$\pi$) closure phase signifying the asymmetric nature of the brightness profiles. R Aqr is the lone symbiotic star with positive asymmetry detected by our AGB survey. In this paper we report followup observations on R Aqr to characterize the observed asymmetry in this symbiotic system. A second objective is to detect a possible water shell around R Aqr and its possible role in the asymmetry detection. The reasons for combining our near-infrared observations with SiO maser observations are the following: (1) probably the water shell emissions detectable in the near-infrared and the SiO emissions detectable in the radio arise from the same circumstellar region \citep{Cotton04, Cotton06}, (2) the SiO maser observations give us some ideas about the distortions from spherical symmetry we might expect in the near-infrared, and also some velocity indications.  

\section{Observations}
\subsection{Infrared Optical Telescope Array (IOTA) Observations}

The observations reported here were carried out at the IOTA 3-telescope array 
equipped with an integrated-optics beam-combiner, IONIC, operating 
in the H-band (1.65 $\mu$m) atmospheric window (\citet{Traub04}, \citet{Berger04}, \citet{Schloerb06}, \citet{Ragland06}) from 30 October to 16 November 2005.

\begin{figure}[bthp]
\centering
\includegraphics[width=0.6\hsize]{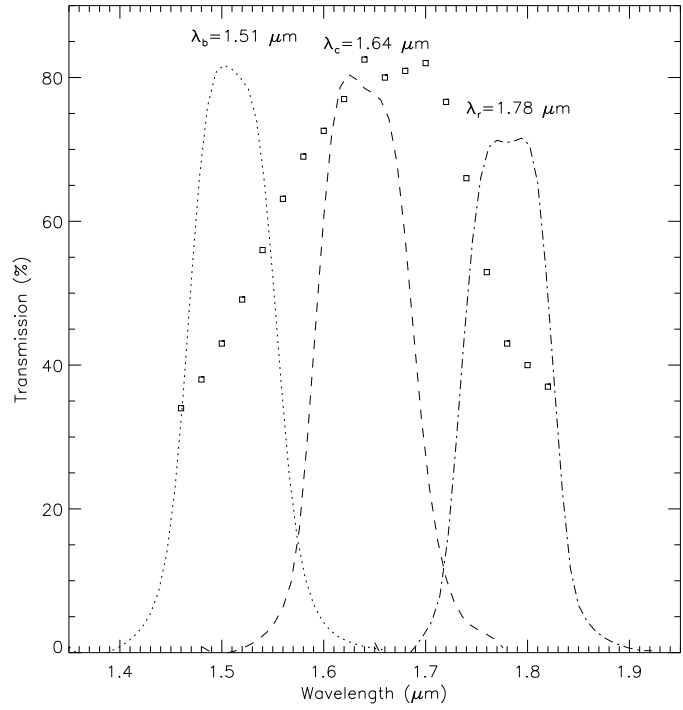}
\caption{The transmission curves of the narrowband filters used in this study are presented here. The three curves from left to right corresponds to the three wavelengths: H$_B$ = 1.51 $\mu$m, H$_C$ = 1.64 $\mu$m \& H$_R$ = 1.78 $\mu$m. Also shown by square symbols is the H band spectrum of R Aqr from \citet{Whitelock07}. 
}
\label{nbFilter}
\end{figure}

The near-infrared spectrum of R Aqr is dominated by water-vapor absorption bands \citep{Whitelock83}. We used three sub--H--band filters to isolate the continuum and absorption-band 
part of the spectrum in order to probe the possible water shell around
R Aqr. Our sub-H-band filters are H$_B$ = 1.51 $\mu$m (water molecular
band on blue side of center), H$_C$ = 1.64 $\mu$m (near-continuum),
and H$_R$ = 1.78 $\mu$m (water molecular band on red side of center); the FWHM values are 0.09, 0.10 \& 0.10 $\mu$m respectively. The bandwidths are a good compromise between sensitivity and spectral purity. The transmission curves for these narrowband filters are shown in Figure~\ref{nbFilter} along with an H band spectrum of R Aqr taken from \citet{Whitelock07} (see also \citet{Whitelock83}). 

\clearpage
\clearpage

Typically, 4 minutes of program star observations were followed by nearby calibrator observations under 
identical instrumental configurations. On each star, we record 4 sets of data
files each containing 200 scans. A scan consists of changing the optical path difference between two beams by roughly 75 $\mu$m in saw-tooth form. We then take 400 scans of shutter data for calibration. The shutter data sequence consists of allowing only one beam at a time (telescope A, B and then C) and at the end blocking all three beams. Each scan takes about 100 ms. The field of view of the instrument in the sense of star light coupled in to the beam-combiner could be estimated as $\sim$ $\lambda$/D = 0.8 arcseconds for D=0.45m, the aperture size of the IOTA telescopes, and $\lambda$ = 1.65 $\times 10^{-6}$m, the wavelength of observations. The instrument field-of-view is further restricted by the fringe scan length to $\sim$ tan$^{-1}(\Delta_{opd}/B_{max})$ = 0.4 arcsec for $B_{max}$ = 26.2995m, the maximum projected baseline of the IOTA observations and $\Delta_{opd}$ = 50 $\times 10^{-6}$m, the smallest fringe scan size of the data block used for bispectral analysis.   

\begin{figure}[bthp]
\centering
\includegraphics[width=0.6\hsize]{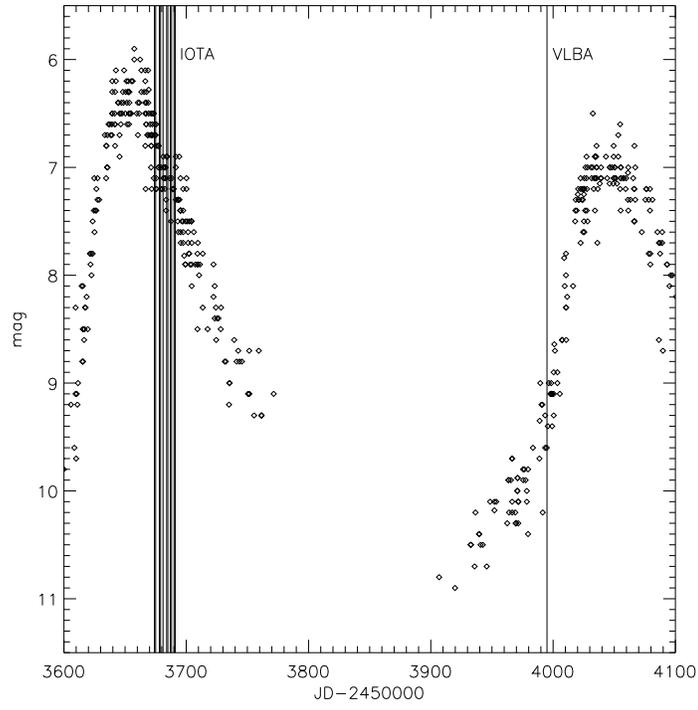}
\caption{The dates of observations from Table 1 (near JD 2443700) and Figures \ref{R_AqrH2O}, \ref{R_Aqr1PCntrVelA} and \ref{R_Aqr2PCntrVelA} are shown on a visual light curve for R Aqr.  The visual data were provided by the AAVSO, www.aavso.org.
}
\label{lightCurve}
\end{figure}

Interferometric fringes were recorded simultaneously on all three baselines of the IOTA enabling three visibilities and one closure phase measurements for each set of fringe packets.
Observations were taken at three IOTA configurations namely, A35B15C10, A25B15C05 and A25B05C05, providing better (u,v) coverage for the science target; where letters A, B and C identify the three telescope apertures and the integers denote the distance of the corresponding aperture in meters from the intersection of South-East and North-East arms. More details of the observations and instrumental configurations are given in Paper I. The pulsation period and phases of R Aqr during our observations are derived from AAVSO and AFOEV databases. A light curve of R Aqr around our interferometric observations is shown in Figure~\ref{lightCurve}.

\clearpage
\clearpage



\subsection{VLBA Observations}
   The observations of R Aqr were made as part of a more extended
VLBA monitoring program of targets being imaged with IOTA and will be
reported on more fully in a separate publication.
The SiO maser observations of R Aqr were made using the NRAO Very Long
Baseline Array (VLBA) on 16 September 2006.
Data were recorded in 2 $\times$ 4 MHz wide channels in each right-- and
left--hand circular polarization centered on the J=1-0,$\nu$=2 and
J=1-0,$\nu$=1  SiO maser transitions at 42.820587 and 43.122027 GHz.
Correlation of the data was performed at the NRAO VLBA correlator in
Socorro, NM, USA and produced 128 channels in all polarization
combinations in each of the two transitions.
Strong continuum sources near each star were observed to serve as
delay, bandpass and polarization calibrators.
R Aqr was observed in 5 observations of 16 minutes duration each.
The quasar J2334+0736 was used as a delay calibrator for R Aqr.

\section{Data Reduction}
\subsection{IOTA Data Reduction}

The recorded interferograms were reduced with an IDL software package. 
Details are given in Paper I. 
In brief, squared visibilities and closure phases are estimated from the three simultaneously recorded fringes through Fourier analysis. The measured fringe power is proportional to the squared visibility and the proportionality constant is calibrated using a nearby calibrator of known visibility. The closure phase is estimated as the phase of the bispectrum of the three fringes. 

The calibrator used for our observations of R Aqr is HD 787. The adopted angular size for this calibrator is 2.43 mas. Two more calibrators, namely, HD 4585 and HD 221745, were observed in order to monitor the stability of the instrument for squared visibility and closure phase measurements. The adopted angular sizes for these calibrators are 1.36 mas and 1.38 mas respectively.
The reduced data are presented in Table~\ref{results} along with baseline information.

\begin{figure}
\epsscale{1}
\plotone{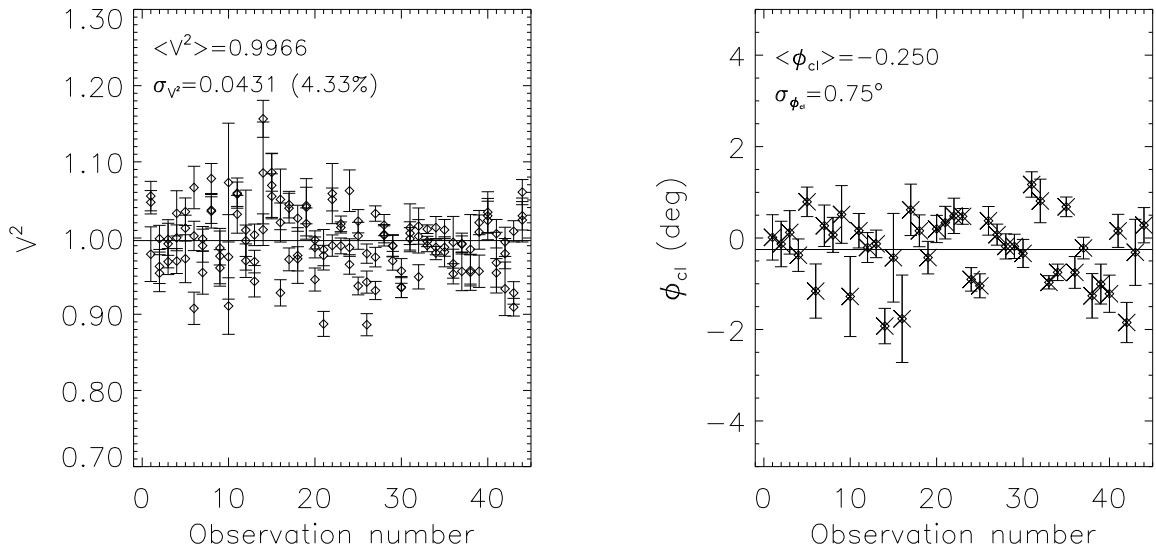}
\caption{Shows the calibrated squared visibility and closure phase measurements 
of several calibrator-pairs observed during 30 Oct. - 18 Nov., 2003, under identical observing configuration 
in order to estimate calibration errors. The estimated calibration error in the squared visibility measurements is $\sim$ 4\% and that of the closure phase measurements is $\sim$ 0.8$^\circ$.
}

\label{calib}
\end{figure}

Typical one-sigma formal errors in the mean values of our uncalibrated closure phase and 
V$^2$ measurements are $\sim$ 0.2$^{\rm o}$ and $\sim$ 1\% respectively.
The formal errors are estimated from the scatters of the 200 fringe scans.  
We estimate calibration errors by observing calibrators under same observing condition 
and calibrating one calibrator with the other after accounting for the finite sizes of both 
calibrators.
The estimated one-sigma calibration error is $\sim$ 4\% for the V$^2$ measurements 
and $\sim$ 0.8$^\circ$ for closure phase measurements (Figure~\ref{calib}). In addition, there is another source of error in the V$^2$ measurements while subtracting background power from the total power in order to estimate the fringe power. Typical value for this additive error is $\sim$ 2\%.
Further discussion on our squared visibility and closure phase measurements can be found in \citet{Ragland04} \& Paper I.

\clearpage
\clearpage


\begin{deluxetable}{ccccccccccccccc}
\tablewidth{0pt}
\tabletypesize{\scriptsize}
\rotate
\tablecaption{Log of R Aqr observations
\label{results}
}

\tablehead{\colhead{} & \colhead{} & \colhead{} & \multicolumn{3}{c}{Baseline BC} && \multicolumn{3}{c}{Baseline AC} 
& &\multicolumn{3}{c}{Baseline AB} & \colhead{} \\
\cline{4-6} \cline{8-10} \cline{12-14}\\
\colhead{Date} & \colhead{$\phi$} & \colhead{$\lambda$ ($\mu m$)} & \colhead{$B_1$ (cm)} &
\colhead{$PA_1$ ($^o$)} & \colhead{$V_1^2$} && \colhead{$B_2$ (cm)} &
\colhead{$PA_2$ ($^o$)} & \colhead{$V_2^2$} && \colhead{$B_3$ (cm)} & \colhead{$PA_3$ ($^o$)} & \colhead{$V_3^2$} & \colhead{$\varphi_{cl}$ ($^o$)}}
\startdata
30 Oct. &0.06&1.51&1479.9& -32.1& 0.248 $\pm$ 0.05&&1627.9&-134.9& 0.173 $\pm$ 0.04&&2431.0&-171.3& 0.040 $\pm$ 0.04&  -4.8$\pm$0.9\\
31 Oct. &0.07&1.51&1160.6&  -3.4& 0.365 $\pm$ 0.05&&2427.4&-127.9& 0.038 $\pm$ 0.04&&3229.3&-145.1& 0.003 $\pm$ 0.04& -17.9$\pm$0.9\\
03 Nov. &0.07&1.51&1317.3& -24.5& 0.310 $\pm$ 0.05&&2014.5&-128.8& 0.080 $\pm$ 0.04&&2665.5&-157.4& 0.019 $\pm$ 0.04& -10.9$\pm$0.9\\
04 Nov. &0.08&1.51&1313.4& -24.2& 0.330 $\pm$ 0.04&&2023.5&-128.7& 0.078 $\pm$ 0.04&&2673.5&-157.1& 0.017 $\pm$ 0.04& -11.4$\pm$0.8\\
10 Nov. &0.09&1.51&1473.7& -44.8& 0.228 $\pm$ 0.04&&1216.8&-137.5& 0.333 $\pm$ 0.04&&1953.8& 173.6& 0.089 $\pm$ 0.04&  -0.9$\pm$1.1\\
12 Nov. &0.10&1.51&1006.3& -30.2& 0.428 $\pm$ 0.05&&1904.2&-127.6& 0.077 $\pm$ 0.04&&2264.8&-153.7& 0.035 $\pm$ 0.04&  -7.8$\pm$0.9\\
13 Nov. &0.10&1.51&1440.2& -44.7& 0.247 $\pm$ 0.04&&1299.9&-134.7& 0.226 $\pm$ 0.04&&1940.7& 177.3& 0.083 $\pm$ 0.04&  -2.1$\pm$0.9\\
16 Nov. &0.11&1.51& 482.1&   2.4& 0.858 $\pm$ 0.07&&1809.1&-127.3& 0.111 $\pm$ 0.04&&2149.2&-137.3& 0.057 $\pm$ 0.04&   2.3$\pm$0.9\\
\hline
30 Oct. &0.06&1.64&1164.0&  -5.1& 0.597 $\pm$ 0.07&&2409.3&-127.7& 0.090 $\pm$ 0.04&&3190.8&-145.6& 0.014 $\pm$ 0.04& -10.0$\pm$0.9\\
31 Oct. &0.07&1.64&1479.9& -32.1& 0.357 $\pm$ 0.05&&1627.9&-134.9& 0.345 $\pm$ 0.04&&2431.1&-171.3& 0.063 $\pm$ 0.04&  -3.8$\pm$0.9\\
04 Nov. &0.08&1.64&1523.0& -33.4& 0.348 $\pm$ 0.05&&1519.9&-137.8& 0.347 $\pm$ 0.04&&2404.4&-175.7& 0.097 $\pm$ 0.04&   0.6$\pm$0.9\\
06 Nov. &0.08&1.64&1315.6& -24.4& 0.369 $\pm$ 0.06&&2018.5&-128.7& 0.163 $\pm$ 0.04&&2669.0&-157.3& 0.038 $\pm$ 0.04&  -9.2$\pm$0.8\\
09 Nov. &0.09&1.64&1170.7& -38.8& 0.470 $\pm$ 0.05&&1744.2&-127.5& 0.296 $\pm$ 0.04&&2078.3&-161.8& 0.138 $\pm$ 0.04&  -2.9$\pm$0.9\\
10 Nov. &0.09&1.64&1316.6& -43.0& 0.423 $\pm$ 0.05&&1541.9&-129.6& 0.319 $\pm$ 0.04&&1966.5&-171.5& 0.171 $\pm$ 0.04&  -0.4$\pm$0.9\\
12 Nov. &0.10&1.64&1307.2& -42.8& 0.423 $\pm$ 0.04&&1557.0&-129.3& 0.321 $\pm$ 0.04&&1971.9&-170.8& 0.151 $\pm$ 0.04&  -3.5$\pm$0.9\\
15 Nov. &0.11&1.64& 499.5& -13.2& 0.834 $\pm$ 0.05&&1507.3&-130.1& 0.336 $\pm$ 0.04&&1789.4&-144.5& 0.219 $\pm$ 0.04&   0.5$\pm$0.8\\
\hline
30 Oct. &0.06&1.78&1289.1& -22.5& 0.453 $\pm$ 0.06&&2079.6&-128.2& 0.139 $\pm$ 0.04&&2727.4&-155.3& 0.037 $\pm$ 0.04& -11.7$\pm$0.8\\
31 Oct. &0.07&1.78&1270.7& -21.0& 0.408 $\pm$ 0.05&&2122.2&-127.9& 0.110 $\pm$ 0.04&&2771.8&-153.9& 0.050 $\pm$ 0.04& -12.2$\pm$0.9\\
03 Nov. &0.07&1.78&1484.8& -32.3& 0.344 $\pm$ 0.05&&1615.9&-135.2& 0.309 $\pm$ 0.04&&2427.4&-171.8& 0.086 $\pm$ 0.04&  -1.0$\pm$0.8\\
04 Nov. &0.08&1.78&1167.6&  -6.5& 0.483 $\pm$ 0.07&&2393.6&-127.6& 0.082 $\pm$ 0.04&&3159.0&-146.0& 0.015 $\pm$ 0.04&  -7.1$\pm$1.1\\
04 Nov. &0.08&1.78&1161.7&  -4.1& 0.437 $\pm$ 0.05&&2420.9&-127.8& 0.073 $\pm$ 0.04&&3215.0&-145.3& 0.012 $\pm$ 0.04&  -7.1$\pm$1.0\\
06 Nov. &0.08&1.78&1169.9&  -7.2& 0.448 $\pm$ 0.06&&2385.1&-127.5& 0.081 $\pm$ 0.04&&3143.0&-146.3& 0.014 $\pm$ 0.04&  -9.3$\pm$0.9\\
10 Nov. &0.09&1.78&1069.8& -34.2& 0.775 $\pm$ 0.09&&1849.6&-127.3& 0.199 $\pm$ 0.04&&2185.9&-156.6& 0.117 $\pm$ 0.04&  -3.6$\pm$0.8\\
15 Nov. &0.11&1.78& 483.9&   4.9& 0.821 $\pm$ 0.07&&1844.9&-127.3& 0.196 $\pm$ 0.04&&2199.4&-136.7& 0.091 $\pm$ 0.04&  -3.5$\pm$0.8\\
\enddata
\end{deluxetable}
\clearpage

\subsection{VLBA Data Reduction}
\begin{figure}
\centerline{
\includegraphics[width=3.5in,height=3.5in]{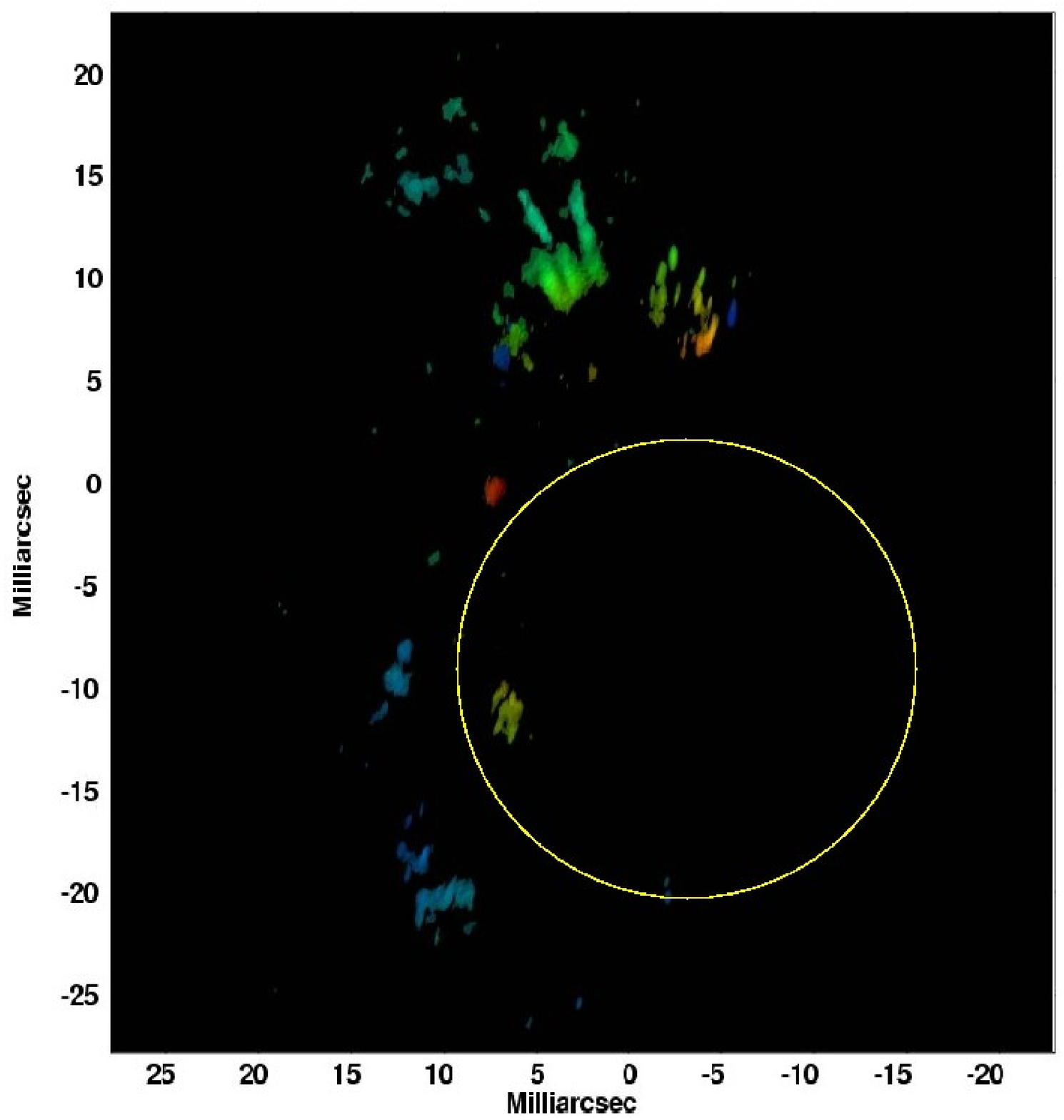}
\includegraphics[width=3.5in,height=3.5in]{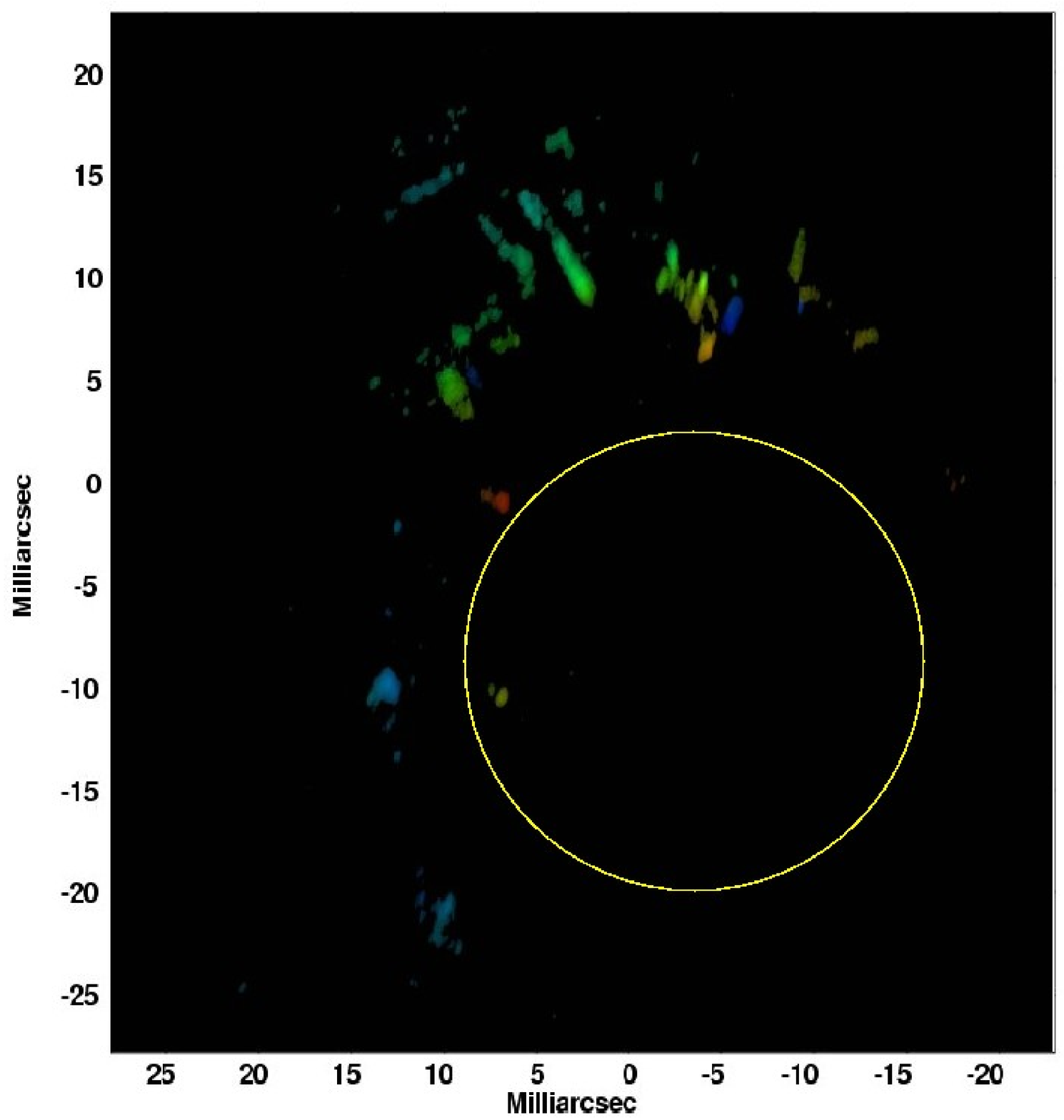}
}
\caption{
{\bf Left:}
SiO J=1-0,$\nu=$2 maser in R Aqr on 16 September 2006.
Color indicates velocity, red=-16.0 km/s, blue=-26.7 km/s, and the
displayed intensity is proportional to the square root of measured intensity.
The yellow circle has the diameter of the  H$_2$O shell from
Table 2 (26.7 mas) and is concentric with the maser ring which is
assumed to be approximately centered on the star.  
\hfill\break
{\bf Right:}
As Left but for J=1-0,$\nu=$1 transition.
}
\label{R_AqrH2O}
\end{figure}

The VLBA SiO maser data were reduced as described in \citet{Cotton04}
and \citet{Cotton06} using a combination of the AIPS and Obit
software packages.
Initial amplitude calibration used measured system temperatures,
weather data and standard antenna gain curves.
Group delay measurements from the continuum calibrator were applied to
the stellar data and a spectroscopic data channel in each transition
was used to self calibrate the phases.
Instrumental polarization was determined from the weakly polarized
continuum calibrators and the linear polarization angle from
comparison with calibration measurements made with the VLA.
Using the phase calibration derived from a single channel, it is then
possible to image each spectral channel and the channel images will be
aligned with each other. Thus, internally consistent images can be derived but the location
relative to the star (not visible in the images) is not measured.
The restricted UV--coverage of these snapshot images coupled with the
large angular extent of the masers required the constrained CLEAN
available using the Obit package task Imager.
The resultant images are shown with velocity coded as color in Figure
\ref{R_AqrH2O}.  
The relative location of the star and the masers was estimated assuming
that the innermost masers lie on a circle or ellipse centered on the
star. 
The resolution of the derived images is represented by an elliptical
Gaussian of 550 $\times$ 160 $\mu$sec with the major axis at position
angle -12$^\circ$.

\clearpage
\clearpage

\section{Discussion}

The reduced infrared data were fitted with various models and the derived parameters are discussed in this section. The parameters of these models are presented in Table~\ref{results2}. We start with a very simple model and progress towards more complex ones. Our new non-zero closure phase measurements presented in this paper are consistent with our earlier finding (Paper I). We will use this information in the more complex models. 

\subsection{Wavelength Dependent Uniform Disk Sizes}

The squared visibility data taken at three wavelengths are fitted separately with a uniform disk (UD) model and the wavelength dependent sizes are derived. We knew from our earlier work (Paper I) that R Aqr has an asymmetric brightness distribution and hence the UD model is not the best one for this object. Nevertheless we started this way in order to (1) compare our results with the UD values available in the literature, (2) investigate the role of possible water shell around R Aqr through wavelength dependent sizes. Moreover, the strength of detected asymmetric features are typically a few \% of the strength of the symmetric component (Paper I), implying that the asymmetric features would only mariginally bias the uniform disk size estimation.

\begin{figure}[bthp]
\centering
\includegraphics[width=0.98\hsize]{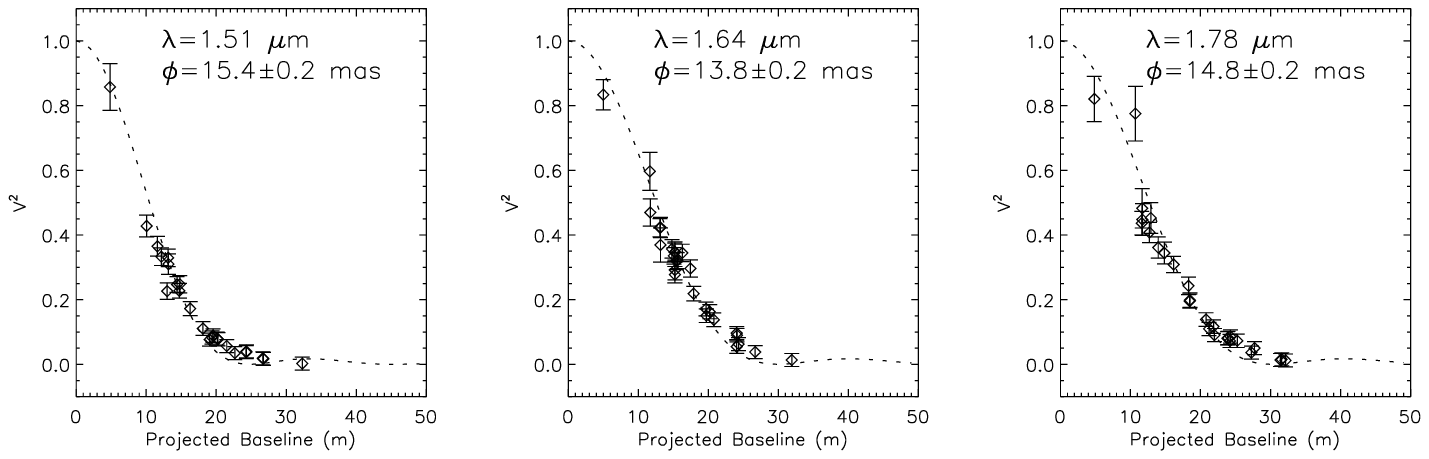}
\includegraphics[width=0.98\hsize]{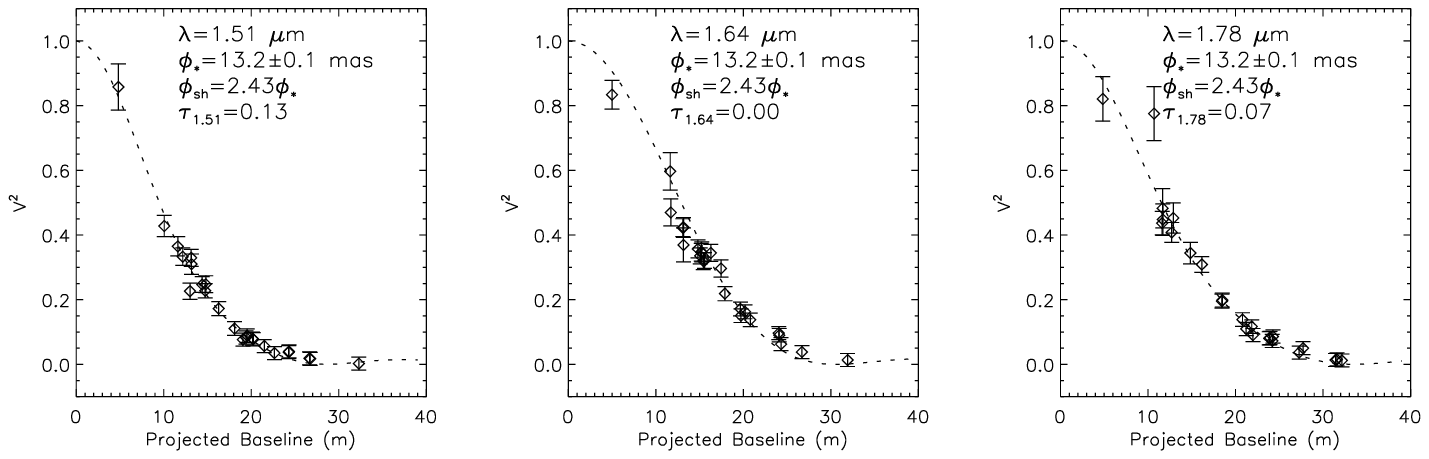}
\includegraphics[width=0.98\hsize]{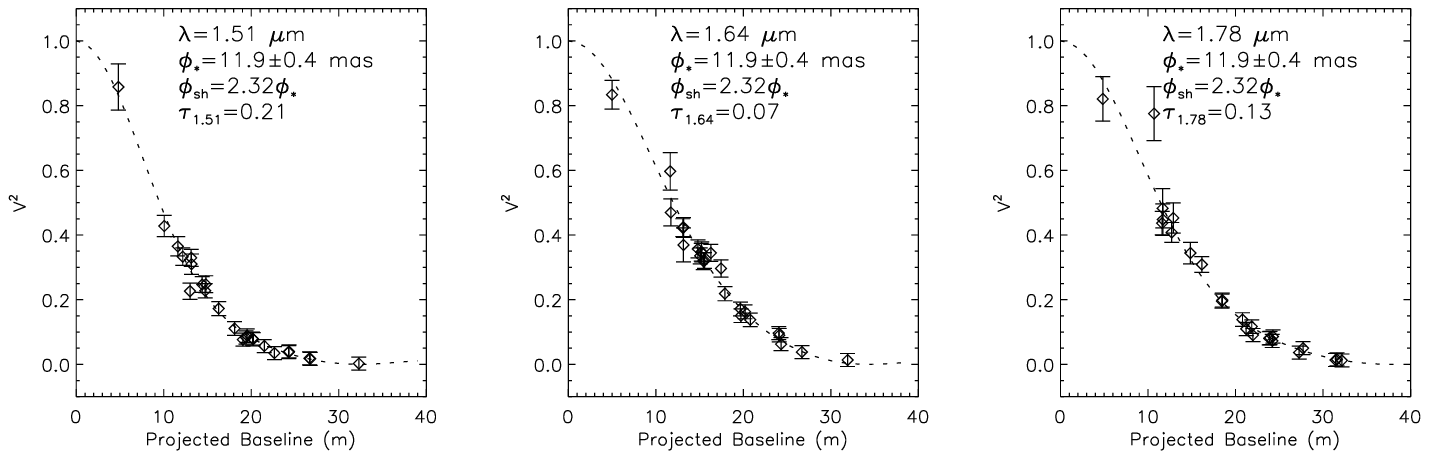}
\caption{
{\bf Top Row:} Squared visibility data points are shown as diamonds along with the error bars. UD models fitted to the data points are shown in dotted lines.
{\bf Middle Row:} Squared visibility data simultaneously fitted with a wavelength-independent UD star surrounded by a water molecular shell. While optical depths at 1.51 and 1.78 $\mu$m are treated as free parameters, the optical depth at 1.64 $\mu$m is fixed at zero. 
{\bf Bottom Row:} Same as middle row except that optical depths at all three wavelengths are treated as free parameters. 
}
\label{diam}
\end{figure}

The derived wavelength dependent angular sizes are shown in Figure~\ref{diam} (Top Row). The derived angular diameters are 15.4$\pm$0.2 mas, 13.8$\pm$0.2 mas and 14.8$\pm$0.2 mas in the 1.51, 1.64  and 1.78 $\mu$m bands respectively. Thus, the size of R Aqr is 12\% and 7\% larger at 1.51 $\mu$m and 1.78$\mu$m repectively than at 1.64$\mu$m. The reduced chi-square value for this model fit to the squared visibility data is 2.6. 

The near-infrared spectrum of R Aqr shows strong absorption bands due
to water molecules in the wings of the near-infrared bands (Figure~\ref{nbFilter}; also\citet{Whitelock83}). Hence, we attribute the apparent wavelength dependent UD sizes to the presence of a water shell around the Mira star. 

\clearpage
\clearpage

\subsection{Wavelength Independent UD stellar size and Water Shell}

We use a two-component model, a wavelength-independent UD star surrounded by a water shell whose optical depth is wavelength dependent \citep{Scholz01}. The model visibility-square and closure phase are estimated through Fourier-Transform of the projected one-dimensional intensity profile. For simplicity, the water shell is assumed to have negligible radial thickness. The sizes of the star and the water shell, and the wavelength dependent optical depths of the water shell are derived through a model fit to the observed multi-wavelength squared visibility data. Assuming that the optical depth of water molecules at 1.64$\mu$m, the pseudo-continuum band, is negligibly small, we fixed this parameter to zero in the model fit. The results are shown in Figure~\ref{diam} (Middle Row). Similar interpretation of a water shell present around the central star can be found in the literature for other Mira stars (\citet{Mennesson02}; \citet{Weiner04}; \citet{Perrin04}).

The free parameters of the model fit are wavelength-independent stellar angular diameter, angular size of the shell and optical depths at 1.51 and 1.78$\mu$m. The optical depths at 1.64$\mu$m is assumed as zero. Stellar effective temperature is constrained from the apparent bolometric magnitude, mbol = 2.23 with an amplitude of variation of
0.91 magnitudes \citep{Whitelock00}, and the angular diameter 
using
\begin{eqnarray}
Log (T_{eff}) = 4.22 - 0.1 m_{bol} - 0.5 log(\phi) \nonumber
\end{eqnarray}
\citep{Ridgway80}. The shell radiative equlibrium temperature is constrained, using
\begin{eqnarray}
T^4_{shell} = \frac{T^4_{eff}}{2} \left[1-\sqrt{1-\left(\frac{\phi_{star}}{\phi_{sh}}\right)^2} +\frac{3}{2}\int^{\infty}_r{\left(\frac{\phi_{star}}{\phi_{sh}}\right)^2\kappa\rho dr}\right]\nonumber
\end{eqnarray}
(\citet{Chandra34}; \citet{Bowen88}). Geometrically thin shell approximation would give,
\begin{eqnarray}
T^4_{shell} = \frac{T^4_{eff}}{2} \left[1-\sqrt{1-\left(\frac{\phi_{star}}{\phi_{sh}}\right)^2} +\frac{3}{2}\left(\frac{\phi_{star}}{\phi_{sh}}\right)^2 \tau_{shell}\right], \nonumber
\end{eqnarray}
where $\tau_{shell}$ is the mean opacity for the shell across the part of the stellar spectrum containing most of stellar luminosity. 

The derived wavelength-independent stellar angular diameter is 13.2$\pm$0.1 mas and the derived shell diameter is $\sim$ 2.43$\phi_*$. This yields an effective temperature of 2737K with an uncertainty due to the bolometric variability of about 11\%. The radiative equilibrium temperature at 2.43 stellar radii, the derived shell position, is 1568K assuming $\tau_{shell}$=0.5. 
This is a reasonable temperature for a high abundance of molecules of CO, CO$_2$, and H$_2$O \citep{Tsuji73}.
The derived optical depths of the water shell are 0.13 and 0.07 at 1.51$\mu$m and 1.78$\mu$m respectively, and the average across the stellar flux should be somewhat higher since we do not observe where these molecules have their peak opacity (i.e. in between our atmospheric windows). The mean optical depth is very unlikely to be below 0.1 or 0.2, corresponding to shell temperatures of 1337 - 1406K.  It is also unlikely to be 1.0 or higher, as then the shell temperature would be $>$ 1761K and the abundances of the molecules would then be lower.  In fact there is probably a self-regulating mechanism operating to keep the optical depth below 1, since higher values lead to higher temperatures that decrease the abundances of the species providing the opacity.


\subsection{Importance of closure phase measurements}
So far, we have used only squared visibility data in our model fits. In this section, we discuss how closure phase measurements could constrain our model fits.

The closure phase measurements of stars (with a fairly sharp edge such
as a uniform disk) have two kinds of high angular resolution
information: (1) A non-zero or non-$\pi$ closure phase signifies an
asymmetric brightness profile (Paper I). 
Asymmetric here means structure which is not rotationally symmetric.
(2) The rough magnitude of closure phase (whether close to zero for a
small star or $\pi$ for a large star) constrains the size of the
central star and thereby could require additional extended
circumstellar components to fit the squared visibility data.    

Centrosymmetric targets should give a closure phase of either zero or $\pm$ $\pi$ depending on how many baselines are beyond the first, second, etc., nulls \citep{Ragland06}. The measured visibilities as a function of baseline and the closure phase data suggest that all measurements reported in this article are taken well within the first null of R Aqr. Hence, the longest baseline data (see Table~\ref{results}), taken on 31 Oct 2005 at 1.51$\mu$m with B$_{max}$=32.293 m that gives a closure phase of 17.9$^{\rm o}$ could provide an upper limit on the possible uniform disk size for the central star. In order for the closure phase measurement to be close to zero rather than $\pi$, the diameter of the central star should be less than 1.22 $\lambda$/B, i.e. the diameter must be $<$ 11.8 mas. Thus, the closure phase data clearly shows that we have overestimated the sizes in the previous two models by at least 13\% with the H$_2$O shell model (Section 4.2) and in the range 17\% to 31\% with UD model (Section 4.1).

Thus, the potential of closure phase measurements is not only in detecting asymmetry but also in
detecting extended symmetric features around the central star. This also suggests that the UD sizes of AGB stars derived from limited squared visibility measurements should be treated with caution. 

\subsection{Improved two component model: Star surrounded by a water molecular shell}

In Figure~\ref{diam} (bottom row), we present a two component model of a central star with wavelength independent size surrounded by a thin shell due to water molecules. The sizes of these two components and the wavelength dependent optical depth are derived through non-linear least-square fit to the observed multi-wavelength squared visibility data. The only difference from the earlier model (Section 4.2) is that the optical depth at 1.64$\mu$m is also treated as a free parameter. Still, we have not used our closure phase data, but the derived size for the central star is now consistent with our upper limit on the stellar diameter from closure phase values (see Section 4.3).


The derived wavelength independent stellar angular diameter is 11.9$\pm$0.2 mas and the derived shell diameter is 2.32$\phi_*$. As before, the stellar effective temperature is derived as 2881 K and that of the shell is estimated as 1691 K assuming $\tau_{shell}$=0.5. The derived optical depths of the shell are 0.21, 0.07 and 0.13 at 1.51$\mu$m, 1.64 $\mu$m and 1.78 $\mu$m respectively. The derived stellar size is consistent with the closure phase measurements. We estimate relatively large optical depths in all three wavelengths. The non-zero optical depth at 1.64$\mu$m could be attributed to a limb darkening effect, scattering of molecules and dust around the star and possible absorption lines in the 1.64$\mu$m filter band. The model over fits the squared visibility data with a reduced chi-square of 0.8.

\subsection{Three component model: Star, Water Shell and an off-axis compact feature}

The next step in our analysis is to account for the non-zero closure phase measurements. As shown in Paper I, the non-zero closure phase could only be explained with an off-center additional component. It could be an off-center (bright or dark) spot on the stellar surface or a companion or dust clump. The simplest possible model would be to add a point source to the existing two-component model.

\begin{figure}[bthp]
\centering
\includegraphics[width=0.98\hsize]{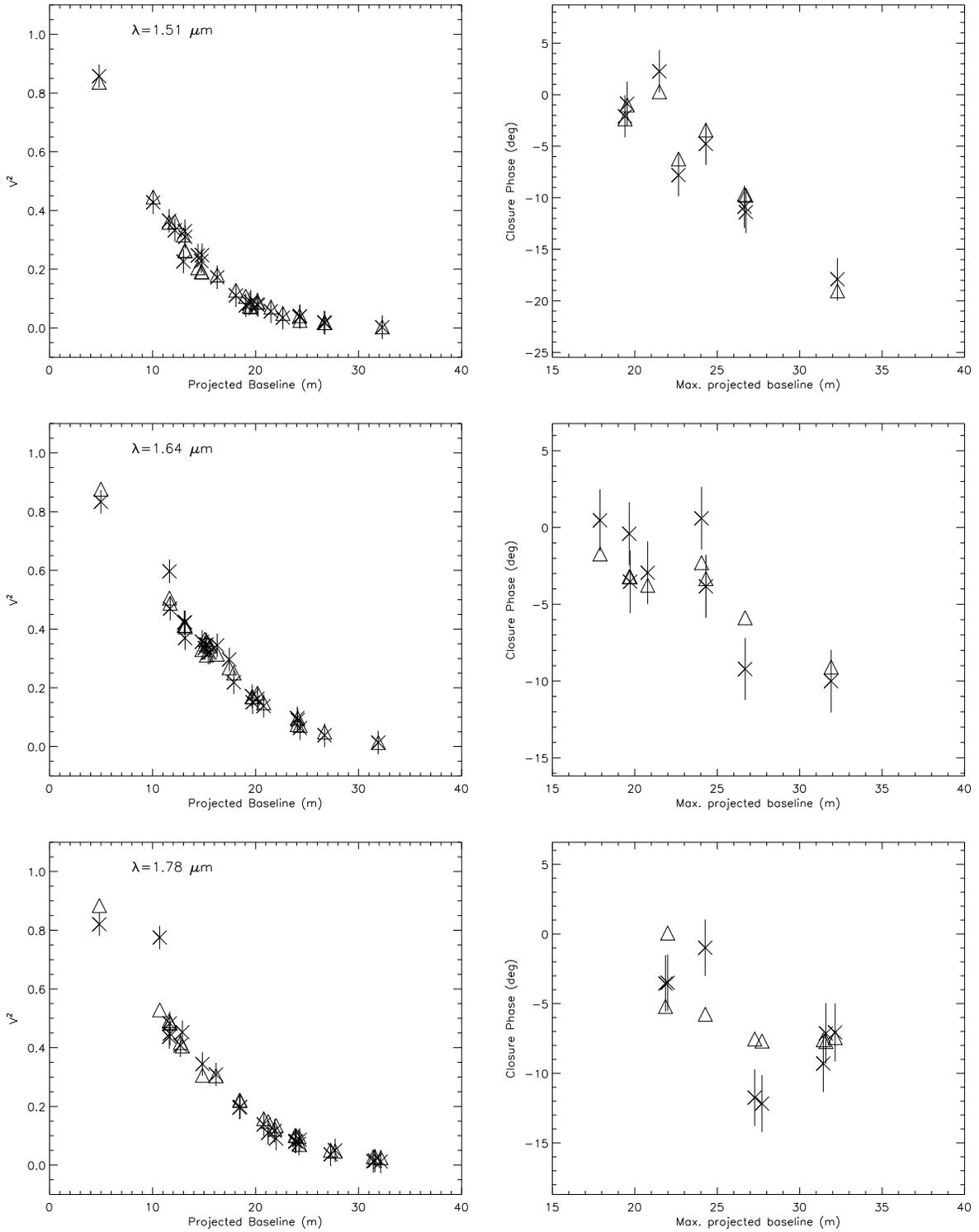}
\caption{Squared visibility and closure phase data fitted with a three-component model: a wavelength independent UD star surrounded by a molecular shell and a off-axis model feature is shown here. The data points along with the error bars are shown in cross symbols and the model values are shown as triangles. 
}
\label{shellBinary}
\end{figure}

Figure~\ref{shellBinary} shows the results of this three component
model. Addition of this third component marginally made changes to the
star and shell parameters. The parameters of this three component model are given in Table~\ref{results2} (Model 4). This three-component model simultaneously fits the squared visibility and closure phase data with a reduced chi-square of 1.15. We also found an alternate solution through Monte Carlo simulation by systematically changing the position of the unresolved compact feature; this model is given in Table~\ref{results2} (Model 5). The derived optical depths of the shell are 0.23, 0.08 and 0.13 at 1.51$\mu$m, 1.64 $\mu$m and 1.78 $\mu$m respectively for Model 4, and 0.26, 0.09 and 0.14 at 1.51$\mu$m, 1.64 $\mu$m and 1.78 $\mu$m respectively for Model 5. The primary difference between the two models (Model 4 \& Model 5) is the position of the off-axis model feature. These derived optical depths are consistent with that reported in the literature for other Mira stars \citep{Perrin04}. 

The derived wavelength independent stellar angular diameter is 11.2 mas and the derived shell diameter is 2.38$\phi_*$. The position angle of the fitted model feature (off-axis point source) is 131$^{\rm o}$ which is roughly orthogonal to the axis of the astrophysical jets detected in the x-ray, UV, visible and radio wavelengths and the angular separation of the off-axis compact feature with respect to the central star is 11 mas. The position angle and angular separation of the off-axis compact feature for the alternate model (Model 5) are 266$^{\rm o}$ and 7.5 mas respectively.

\citet{Tuthill00} observed R Aqr at near-infrared wavelengths using
aperture masking interferometry at the Keck-I telescope with an angular resolution of $\sim$ 15 mas. These authors did not detect any companion with a relative brightness of $\le$ 5 magnitudes.   
This is probably consistent with any of the models in Table 5, since a
2--5\% feature at a separation of 0.5-0.7 times the angular resolution
is not likely to have been detected.

\clearpage
\clearpage

\begin{deluxetable}{llcllclllrr}
\tablewidth{0pt}
\tabletypesize{\scriptsize}
\rotate
\tablecaption{Results of the four model fits.
\label{results2}
}
\tablehead{
& $\phi_{\lambda}$ &&
\multicolumn{2}{c}{H$_2$O shell}&&
\multicolumn{3}{c}{Off-axis Feature} &&\\
\cline{4-5} \cline{7-9}                   
\colhead{Model} &\colhead{(mas)}&&\colhead{$\phi_{sh}(mas)$} & \colhead{Optical Depth}&&
\colhead{$\delta\phi$(mas)} & \colhead{PA} &  \colhead{Flux ratio} & \colhead{$\chi^2_R$(V$^2$)} & \colhead{$\chi^2_R$(total)}
}
\startdata
&&&&&&&&&&\\
1) Star (UD)&$\phi_{1.51}$=15.4$\pm$0.2&&-&-&&-&-&-&&\\
&$\phi_{1.64}$=13.8$\pm$0.2&&-&-&&-&-&-&&\\
&$\phi_{1.78}$=14.8$\pm$0.2&&-&-&&-&-&-&2.6&$^1$2481\\
&&&&&&&&&&\\
\hline
&&&&&&&&&&\\
2) Star + Shell &$\phi_{1.51}$=13.2$\pm$0.1&&32.0$\pm$2.6&$\tau_{1.51}$=0.13$\pm$0.04&&-&-&-&&\\
($\tau_{1.64}$ = 0) &$\phi_{1.64}$=13.2$\pm$0.1&&"&$\tau_{1.64}$=0.00&&-&-&-&&\\
&$\phi_{1.78}$=13.2$\pm$0.1&&"&$\tau_{1.78}$=0.07$\pm$0.04&&-&-&-&1.3&$^1$1827\\
&&&&&&&&&&\\
\hline
&&&&&&&&&&\\
3) Star + Shell&$\phi_{1.51}$=11.9$\pm$0.2&&27.6$\pm$1.9&$\tau_{1.51}$=0.21$\pm$0.05&&-&-&-&&\\
&$\phi_{1.64}$=11.9$\pm$0.2&&"&$\tau_{1.64}$=0.07$\pm$0.04&&-&-&-&&\\
&$\phi_{1.78}$=11.9$\pm$0.2&&"&$\tau_{1.78}$=0.13$\pm$0.05&&-&-&-&0.8&$^1$21\\
&&&&&&&&&&\\
\hline
&&&&&&&&&&\\
4) Star + Shell +&$\phi_{1.51}$=11.2$\pm$0.8&&26.7$\pm$0.8&$\tau_{1.51}$=0.23$\pm$0.05&&11$\pm$4&131$\pm$17$^{\rm o}$&R$_{1.51}$=2.2$\pm$0.3\%&&\\
Off-axis&$\phi_{1.64}$=11.2$\pm$0.8&&"&$\tau_{1.64}$=0.08$\pm$0.02&&"&"&R$_{1.64}$=2.4$\pm$0.4\%&&\\
feature&$\phi_{1.78}$=11.2$\pm$0.8&&"&$\tau_{1.78}$=0.13$\pm$0.04&&"&"&R$_{1.78}$=3.1$\pm$0.4\%&1.02&1.15\\
&&&&&&&&&&\\
\hline
&&&&&&&&&&\\
5) Star + Shell (Alternate solution) +&$\phi_{1.51}$=11.2$\pm$0.8&&26.6$\pm$0.7&$\tau_{1.51}$=0.26$\pm$0.06&&7.5$\pm$0.7&266$\pm$7$^{\rm o}$&R$_{1.51}$=3.0$\pm$0.5\%&&\\
Off-axis&$\phi_{1.64}$=11.2$\pm$0.8&&"&$\tau_{1.64}$=0.09$\pm$0.02&&"&"&R$_{1.64}$=4.0$\pm$0.5\%&&\\
feature&$\phi_{1.78}$=11.2$\pm$0.8&&"&$\tau_{1.78}$=0.14$\pm$0.04&&"&"&R$_{1.78}$=5.5$\pm$0.5\%&1.12&1.10\\
&&&&&&&&&&\\

\enddata

\tablenotetext{1} {These values are not used for chi-square optimization (see text).\\
$\phi_{\lambda}$ and $\phi_{sh}$ are the angular diameters of the central star and the molecular shell respectively. $\delta\phi$ is the angular separation between the central star and off-axis model feature.} 
\end{deluxetable}
\clearpage


\subsection{Accretion Disk Hypothesis}

In this section we focus on whether or not the fitted model feature in the IOTA observations
can be accounted for by the white dwarf companion that has been previously discussed in
the literature (e.g.,\citet{Hollis97}) or by the effect of Roche lobe hot spot.  First, we discuss the luminosity ratio between the white dwarf
companion and the AGB star.  Next, we calculate the effect of the accretion disk to determine its 
luminosity contribution (4.6.2), and a number of other relevant physical effects including mass loss 
on the accretion luminosity (4.6.3), heating of the AGB star from the white dwarf (4.6.5), heating of the accretion 
disk from the AGB star (4.6.6), and the luminosity of the hot spot of the accretion disk (4.6.7).   Understanding 
the magnitude of the infrared emission due to these physical effects allows us to determine 
whether or not the 
companion feature is consistent with a white dwarf star with the orbital parameters in the literature,
or if it could be an additional white dwarf companion, or if the companion feature is really due
to a physical effect of the AGB star itself, such as a hot spot near the edge of the star. 

\subsubsection{White dwarf Luminosity}


We begin this discussion with a simple calculation of the luminosity of the white dwarf
companion compared to that of the red giant based on parameters for these stars in the literature. We assume the red giant star has a radius of 300 R$_{\odot}$, a temperature of 2800 K, giving it a total luminosity of 5$\times$10$^3$ L$_{\odot}$ \citep{Kafatos86, Hollis00b, Burgarella92, Makinen04}. We adopt a distance of 200 pc based on the work of \citet{Hollis97}. 

Two different sets of parameters for the companion have been used in the literature. One choice is that the companion has a temperature of around 40,000 K and a diameter of 0.1 R$_{\odot}$,
which would be classified as a subdwarf star, with a total luminosity of approximately 10 L$_{\odot}$. The other choice is a white dwarf star, with a temperature of about 130,000 K, and a diameter of 0.014 R$_{\odot}$, also giving approximately 10 L$_{\odot}$ for the total luminosity. 

We begin by calculating the ratio of the luminosity at 1.65 $\mu$m of the companion to that of the red giant star. The result is that the ratio for the subdwarf case is about 10$^{-5}$, whereas for the white dwarf the ratio is 7.1$\times$10$^{-7}$, which is substantially smaller. Given that the ratio of
luminosities for the observed asymmetry corresponds to about 2$\%$ of the total 1.65 $\mu$m
luminosity of the red giant star, it is clear that the asymmetry cannot be due to the detection of the companion itself. Hence the accretion disk around the white dwarf star or subdwarf companion must contribute the bulk of the luminosity at this wavelength to explain the observational results. For the remainder of this section we explore a number of physical effects associated with the companion and its accretion disk that could explain the observations.

\subsubsection{Accretion Luminosity}

The accretion luminosity can be readily calculated based on the results of \citet{Lynden74} and \citet{Pringle81}. We now calculate the luminosities at 1.65 $\mu$m for the two cases. For an optically thick, physically thin, flat disk, the luminosity at the earth is given by;

\begin{equation}
L_{acc}\left(\nu\right) = \frac{\int_{R_*}^{R_{out}} 2\pi R\pi B_{\nu}[T(R)] dR}{D^2}
\end{equation} 
where

\begin{equation}
T\left(R\right) = \left(\frac{3GM_*\dot{M}}{8\pi\sigma} \left[1-\left(\frac{R_*}{R}\right)^{1/2}\right]\right)^{1/4}
\end{equation}
is the temperature in the accretion disk \citep{Pringle81} and D is the distance to the Earth. The quantity B$_{\nu}$ is the Planck function, M$_*$ is the mass of the white dwarf, R$_*$ is its radius, $\dot{M}$ is the accretion rate onto the disk, and $\sigma$ and G are the Stefan-Boltzmann constant and the gravitational constant, respectively. The maximum temperature in the accretion disk occurs at a radius R$_{max}$ = 1.36 R$_*$, and is given by

\begin{equation}
T_{max} = 0.488 \left(\frac{3GM_*\dot{M}}{8\pi \sigma R^3_*}\right)^{1/4}
\end{equation}
as discussed in \citet{Lynden74}, \citet{Pringle81}, and \citet{Hartmann98}. In this treatment we neglect the emission from the boundary layer, which occurs over a very small angular region around the surface of the white dwarf star and emits at a very high temperature, which would have a negligible contribution to the luminosity in the infrared region near 1.65 $\mu$m. 

With the parameters for the white dwarf star and the AGB star as discussed previously in this section, we can calculate the ratio of the AGB luminosity to that of the accretion disk. The red giant luminosity is given by:

\begin{equation}
L_{AGB}(\nu)= \pi\left(\frac{R_*}{D}\right)^2 B_{\nu}\left(T_*\right)
\end{equation}
where R$_*$, T$_*$ are the radius and temperature of the star, and D is the distance to the Earth. 

From these equations we calculate the ratio of the accretion luminosity at 1.65 $\mu$m to that of the AGB star, assuming different values for the mass loss rate from the AGB star and an adopted efficiency of capture of material onto the accretion disk of 10\%. Thus, if the mass loss rate were 10$^{-5}$ M$_{\odot}$ yr$^{-1}$, then the accretion rate would be 10$^{-6}$ M$_{\odot}$ yr$^{-1}$. Table \ref{lumCal} presents our
results for varying mass loss rates. We find that the maximum contribution to the 1.65 $\mu$m flux from the accretion disk occurs at the highest conceivable mass loss rate from the AGB star, namely at 10$^{-5}$ M$_{\odot}$ yr$^{-1}$, and the ratio is not strongly dependent on the assumptions about the white dwarf star. The ratio of 1.6\% (Table 3) is reasonably close to that of the closure phase measurements for the expected intensity ratio, so we must determine whether the mass loss rates are very high, as required to explain the measurements, or whether they are relatively low.

\clearpage
\begin{deluxetable}{cccccc}
\tablewidth{0pt}
\tabletypesize{\scriptsize}
\tablecaption{Accretion luminosities of White Dwarf companion to AGB star at 1.65 $\mu$m.
\label{lumCal}
}

\tablehead{\colhead{WD} & \colhead{WD} & \colhead{AGB Mass} & 
\colhead{WD} & \colhead{T$_{acc,max}$}&\colhead{L$_{AGB}$/L$_{Accretion}$}\\
\colhead{Radius}&\colhead{Temperature}&\colhead{Loss Rate}&\colhead{Acc. rate}&\colhead{(K)}&\\
\colhead{(R$_{Sun}$)}&\colhead{(K)}&\colhead{(M$_{Sun}$ yr$^{-1}$)}&\colhead{Rate$^1$}&&\\
}
\startdata
&&&&&\\
0.1  & 4.0$\times$ 10$^4$& 10$^{-5}$&10$^{-6}$&4.8$\times$ 10$^4$&1.5$\times$ 10$^{-2}$\\
     &                   & 10$^{-6}$&10$^{-7}$&2.3$\times$ 10$^4$&3.0$\times$ 10$^{-3}$\\
     &                   & 10$^{-7}$&10$^{-8}$&1.3$\times$ 10$^4$&5.7$\times$ 10$^{-4}$\\
0.014& 1.3$\times$ 10$^5$& 10$^{-5}$&10$^{-6}$&1.8$\times$ 10$^5$&1.6$\times$ 10$^{-2}$\\
     &                   & 10$^{-6}$&10$^{-7}$&1.0$\times$ 10$^5$&3.4$\times$ 10$^{-3}$\\
     &                   & 10$^{-7}$&10$^{-8}$&5.6$\times$ 10$^4$&7.1$\times$ 10$^{-4}$\\
\enddata

\tablenotetext{1} {WD accretion rate is assumed to be 10\% of AGB Mass Loss Rate.}
\end{deluxetable}
\clearpage

\subsubsection{Mass loss estimates}

A variety of mass loss rates have been estimated for R Aqr. The largest rate is 8.9$\times$10$^{-7}$ M$_{\odot}$ yr$^{-1}$ (Danchi et al. 1994), while \citet{Spergel83} and Hollis et al. (1985) have much lower rates, 1.9$\times$10$^{-8}$ M$_{\odot}$ yr$^{-1}$ and 4.2$\times$10$^{-8}$ M$_{\odot}$ yr$^{-1}$, respectively. These measurements were
made using a variety of techniques, for example Danchi et al. (1994) used infrared
interferometric measurements of the distribution within a few stellar radii of the star itself,
while \citet{Spergel83} used measurements of line profiles of CO, to infer mass loss rates.

The largest of these gives a luminosity ratio of 0.3\%, which is too small to account for the observed closure 
phases.  To obtain accretion luminosities large enough to account for the observations, a much larger fraction of the AGB wind (than 10\%)  must have been captured by the white dwarf star than was used in the previous estimate.
Alternatively,  the more recent measurements were taken during a period of substantially larger mass loss than were used in our estimates in the previous section.

Both alternatives are possible, as Danchi et al. (1994) and Bester et al. (1996) have shown
that mass loss can be episodic and that there can be periods of increased mass loss relative to the
mean mass loss rate, that would be obtained from CO line profile measurements, which probe a
much larger volume around a star than the interferometric measurements. Thus, the CO
measurements probe a longer history (of the order of 1000 years) and hence average over periods when the mass loss could have been relatively low.

\subsubsection{Orbital parameters}

It is also possible that the fraction of the AGB wind captured could vary substantially in
time, and this is due to the fact that the orbit is eccentric. Figure~\ref{geometry}  displays the orbit of the white
dwarf companion to the AGB star over the course of a complete orbital period, P=44 years,
beginning in 1974, the date of periastron, at T=2442100.0 JD. The parameters used in the
calculation are from Hollis, Pedelty, and Lyon (1997) and they are: a, the semi-major axis,
between 16 and 18 AU for a total system mass between 2.5 and 3 M$_{\odot}$; the eccentricity, e=0.8; the inclination, i $\sim$ 70
degrees (in their paper); the longitude of the ascending node, $\Omega$ $\sim$ 90 degrees; and the longitude of periastron,
$\omega$ $\sim$ $\pm$ 90 degrees. These orbital parameters do not produce an orbit that matched the position angle reported in \citet{Hollis97}. Instead we find that we can match the reported position angle if we use $\omega$ = -90 degrees and $\Omega$ = -75 degrees. Given the large uncertainty in the parameters in \citet{Hollis97}, the adopted parameters are reasonable and consistent with all the data discussed herein (In the absence of radial velocity measurements, the inclination angle is subject to an ambiguity. Accordingly, the inclination, i $\sim$ 110 degrees; the longitude of the ascending node, $\Omega$ $\sim$ +105 degrees; and the longitude of periastron, $\omega$ $\sim$ +90 degrees is an alternate orbital solution for the R Aqr binary system). The plot (Figure~\ref{geometry}) displays the orbit (solid line), location at four year intervals as
labeled, and the AGB star (gray filled circle at the origin) and the white dwarf accretion disk
(outline in black) foreshortened due to the high inclination of the orbit, assuming the accretion
disk and the stars orbit in the same plane. The AGB star and accretion disk are also to scale,
where we have adopted a distance of 200 pc. It's worth mentioning that the orbital parameters derived by Hollis et al. 1997 employed a
number of assumptions based on largely circumstantial evidence and a
single measurement of the offsets between the centroids of the SiO
masers surrounding the Mira and the continuum HII region thought to be
surrounding the white dwarf companion.  While the centroids of these
features may not be coincident with the star responsible for them,
there is a good case made that in 1996, the companion white dwarf
appeared several 10's of mas north of the Mira.  Unless the orbital
period is much shorter than the assumed 44 years, the object detected
by Hollis et al. 1997 is unlikely to be the fitted model feature (Model 4 \& Model 5). 

\citet{McIntosh07} analyzed radial velocity data obtained from 1946 until 2007 from a 
SiO maser emission, near-infrared and visual spectral lines. They determined an orbital period 
of 34.6 $\pm$ 1.2 yr, an eccentricity of 0.52 $\pm$ 0.08, and a projected semi-major axis of 
3.5 $\pm$ 0.4 AU. Their results indicated periastron passage occurred in 1980.3 $\pm$ 0.8 and 
the angle between the line of nodes and periastron is 110.7 $\pm$ 18.4 degrees. These parameter 
values would indicate Roche lobe overflow at periastron, however, these results are suspect as 
they produce an unphysical system mass of 0.043 M$_{\odot}$, which disagrees with the expected 
system mass of 2.5 to 3.0 M$_{\odot}$ as discussed previously. \citet{McIntosh07} can obtain a 
somewhat higher system mass if they use a much smaller inclination angle of 20 degrees, rather 
than $\sim$ 70 degrees, as expected from other analyses. Continued monitoring and analyses of 
imaging and spectral data is needed to produce more precise estimates of the orbital elements 
for this system. 

%

\subsubsection{Heating of primary from secondary}

There are a number of other effects that are worth exploring, given knowledge of the
orbit of the two stars. One type of effect that has been noted in the literature on cataclysmic
variable stars is a warming effect on the primary star due to radiation from the secondary star.
For example, it is possible that the white dwarf star, which is illuminating the AGB star, could
potentially heat up one side of the star, and create an asymmetry, which could be observable in
the closure phases at 1.65 $\mu$m. Based on the work of Warner (1995), it is easy to show that the
fractional temperature change of the AGB star, $\Delta$T/T, is:

\begin{equation}
\frac{\Delta T}{T} = \frac{\pi}{8} \left(\frac{R_2}{d}\right)^2 \left(\frac{L_1}{L_2}\right)
\end{equation}
where R$_2$ and L$_2$ are the radius and luminosity of the AGB star, respectively, and L$_1$ is the luminosity of the white dwarf star, and d is the distance between them. At periastron, Eqn. (5)
gives a fractional temperature change of 2$\times$10$^{-5}$, which is a temperature change of 0.056 K, which
is a very small effect indeed. At the distance in 2004, d=26 AU, this effect is completely
negligible.

\subsubsection{Heating of secondary from primary}

However, as the white dwarf star heats the AGB star, the AGB star likewise heats the
white dwarf star itself and its accretion disk. Given the large luminosity of the AGB star, $\sim$ 5000
L$_{\odot}$, the effect of the AGB star on the accretion disk, particularly its cooler, outer parts, could be
substantial. To understand the size of this effect, we make a simple calculation of a black dust
grain in the presence of this luminosity:

\begin{equation}
T_{grain} = \left[\frac{1}{4}\right]^{1/4} \left(\frac{R_{star}}{d}\right)^{1/2} T_{star}
\end{equation}
where R$_{star}$ and T$_{star}$ are the radius and temperature of the star and d is the distance from the
star to the dust grain.
We assume the AGB star has a temperature of
2800 K and a radius of 300 R$_{\odot}$. For a grain at 5.3 AU, we obtain a temperature of 1016 K, while
at larger distances, such as 17 AU, the temperature is substantially lower, about 567 K, and at the
location of the white dwarf in 2004, 26 AU, the temperature is 458 K.

We can estimate the increase in luminosity of the accretion disk due to the warming
effect from the AGB star. Using eqn. (1) and assuming a constant temperature from about 0.5
of the outer radius of the accretion disk (about 1 AU), to the outer radius, we find the luminosity
at 1.65 $\mu$m is 20 Jy or 1.3\% at 5.3 AU, and 0.6 mJy or 4$\times$10$^{-5}$ \% at a distance of 26 AU. For the
current epoch, the contribution to the 1.65 $\mu$m luminosity is negligible, but there will clearly be a
strong effect as the stars approach periastron, where the distance is a(1-e)=3.4 AU, where we
have used the orbital parameters for the lower total mass. In this case, the effect is of the order
of 113 Jy or 7\% of the 1.65 $\mu$m flux density of the AGB star, so this should be observable with
current and future long baseline interferometers in the next few years.

\subsubsection{Effect of Roche lobe hot spot}
Finally, another effect that could potentially be of importance observationally as the two
stars approach periastron is the emission from a bright spot caused by Roche lobe overflow. A simple estimate of the flux
produced by the bright spot is given in Warner (1995), pp. 81-82. The flux, which we write as,
F$_{bs}$($\nu$), is

\begin{equation}
F_{bs}(\nu) = \left(\frac{\Delta R_s}{d}\right) \left(\frac{R_d}{d}\right) \Delta \theta ~B_{\nu}(T_{bs})\left(\frac{d}{D}\right)^2
\end{equation}
where $\Delta R_s$ the radial size of the spot, $\Delta \theta$ is the radial angle along the accretion disk taken up by
the spot, R$_d$ is the outer radius of the accretion disk. The symbols d and D are the distance
between the two stars and the distance to the observer, respectively. At periastron, we find the
hot spot is expected to contribute as much as 110 Jy (or 7\%) at 1.65 $\mu$m, assuming a typical hot spot
temperature of $\sim$ 10000 K.
%

\begin{figure}[bthp]
\centering
\includegraphics[width=0.6\hsize]{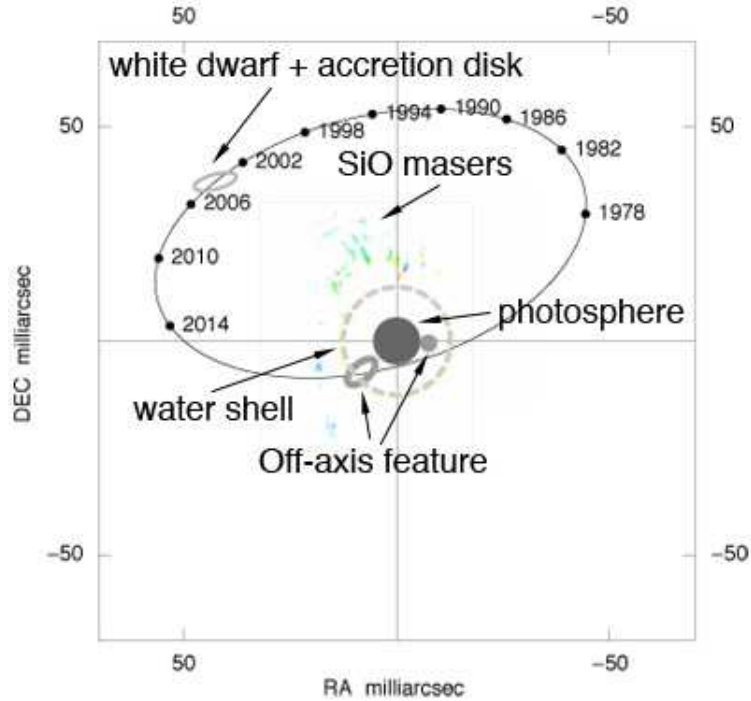}
\caption{
Orbit of the white dwarf companion to R Aqr based on the orbital elements published
by Hollis, Pedelty, and Lyon (1997) with the longitude of the ascending node and periastron modified as discussed in Section 4.6.4, computed for the epoch of the IOTA observations. 
The Mira star, H$_2$O shell and the fitted model feature (at the two possible locations from Table 2) 
are also shown here. The SiO maser emission in R Aqr on 16 September 2006 is superimposed on to this figure where color indicates velocity, red=-16.0 km/s, blue=-26.7 km/s.
}
\label{geometry}
\end{figure}

We conclude that a hotspot from Roche lobe overflow is energetically capable of producing a feature that could, then, explain the observed closure phases.  However, Roche lobe overflow is not expected for the orbit shown in Figure~\ref{geometry}, as the separation is much too large. 


\clearpage
\clearpage

\subsection{Clumpy water shell}
So far we have assumed that the central star and the water shell are symmetric in nature and attributed the observed asymmetric feature to an off-axis point source (accretion disk around white-dwarf companion or
a bright hot spot). In this section, we explore the possibility of an
asymmetric water shell (without a need for a close companion or hot
spot) in explaining the observed asymmetry. 
We replaced the model fitted off-axis point sources (section 4.5) with Gaussian-shaped marginally-resolved features and found no significant degradation in the reduced chi-square. This signifies that a clumpy molecular shell would probably fit the measured data. In the absence of any a-priori knowledge on the shape of the asymmetric shell, our obvious choice was to go for a model independent image reconstruction algorithm to explore this possibility further. The limited uv coverage of our observations (Figure~\ref{uv}) would prevent reconstruction of the entire instrument field of view of $\sim$ 0.4 arcsecond. However, our Monte Carlo simulations looking for an unresolved feature around R Aqr (Section 4.5) did not detect any feature in the annular region with an inner radius of $\sim$ 20 mas and the outer radius of $\sim$ 0.4 arcsecond. Hence, we performed the image reconstruction of the central $\sim$ 20 mas area around the central star. 

\subsubsection{Direct minimization algorithm for image reconstruction}
The van Cittert Zernike Theorem states that the spatial coherence
function is the Fourier transform of the brightness distribution of the astronomical source under certain conditions. Thus, in principle, an inverse Fourier transform of spatial coherence function should provide the brightness distribution of the target. Long baseline interferometeters measure only fringe visibility amplitudes (but not fringe phases) as a function of baselines and closure phases for all closed triangles. Hence, lack of fringe phase information challenges this simple Fourier inversion technique to reconstruct the images. 

Image reconstruction from measured visibility and closure phase data
is a well known procedure in the context of speckle interferometry
\citep{Weigelt91}. The challenges have been in adapting this method to optical interferometric arrays with sparse apertures. There is no unique way of expanding the limited baseline data to all of the bispectral domain or its subdomain where the unique image reconstruction of an arbitrary  object is possible.
However, the positivity constraint could be used to recover the image of relatively compact objects.

The standard radio interferometry self calibration techniques assume
direct measurement of the complex visibilities whereas the much
shorter atmospheric coherence times of Optical/IR interferometers
requires different techniques for estimating visibility squared from a
power spectrum analysis and closure phase from a bispectrum analysis.
This means that the usage of radio techniques are difficult to apply
to our data.
Therefore we developed an algorithm
based on direct minimization procedure where the model image plane  
intensity is varied to reduce the function
\begin{equation}
F=\alpha \sum_l\left(\frac{V^2_{l}-\bar{V^2_{l}}}{\sigma_{V^2_l}}\right)^2+\sum_{{i,j,k}}\left(\frac{B_{ijk}-\bar{B}_{ijk}}{\sigma_{B_{ijk}}}\right)^2,
\label{E1}
\end{equation}
under the following constraints
\begin{eqnarray}
f(\bf{r})& \ge 0, & {\rm if}~~ {\rm r \le r_0} \nonumber\\
f(\bf{r})& =0, & {\rm otherwise,}
\label{E2}
\end{eqnarray}
where $f(\bf{r})$ is the object brightness distribution and $\bar{B}_{ijk} = \bar{V}_i \bar{V}_j \bar{V}_k
exp[i(\bar{\varphi}_{i}+\bar{\varphi}_{j}+\bar{\varphi}_{k})]$ represents bispectral elements of the reconstructed image. The Fourier component of the reconstructed image for l-th baseline is written as
$\bar{V}_l exp(i\bar{\varphi}_{l})$.
The summation is performed for all measured visibilities $V_l$ and
bispectral elements $V_i V_j V_k exp(i\varphi_{ijk})$ where $\varphi_{ijk}$
are measured closure phases ($\bar{\varphi}_{i}+\bar{\varphi}_{j}+\bar{\varphi}_{k}$).
The measured visibilities and bispectral elements are weighted according
to their errors $\sigma_{V^2_l}$  and $\sigma_{B_{ijk}}$.
The relative contribution of pure visibilities and phase related values is balanced by using the constant $\alpha$.
The parameter $r_0$ defines the field of view of the interferometer. 



The minimization procedure enforces positivity constraints by using the multipliers method \citet{Bertsekas82}.
In this minimization procedure, the constraints (Equation~\ref{E2}) are combined with the objective function (Equation~\ref{E1}) to form a modified Lagrangian function of the form

\begin{equation}
L_c=F+\frac{1}{2c}\sum_j ({[max(0,\gamma_j-c f({\bf r}_j))]^2-\gamma_j^2})
\label{E3}
\end{equation}
and the minimization problem (Equations~\ref{E1} \&~\ref{E2}) is replaced with a sequence of unconstrained minimization
procedures of the form,
\begin{eqnarray}
{\rm minimize} & L_{c_k}(f^k,\gamma) 
\end{eqnarray}
where $\{c_k\}$ is a growing positive sequence of penalty parameter $c_k$, and minimization is performed for all n restored points of the image inside the FOV.
And, $f^k({\bf r})$ is the real-valued solution that provides the minimum of the modified Lagrange function $L_{c_k}$ which is used as the first approximation for the minimization of $L_{c_{k+1}}(f,\gamma)$.

The new penalty parameter for the next minimization step is computed as
\begin{equation}
c_{k+1}=(1+\beta) c_k,
\end{equation}
where $\beta$ is a constant $>$ 0.

For each new penalty parameter $c_{k+1}$ the Lagrange vector  ${\gamma_k}$
is recalculated as
\begin{equation}
\gamma^{k+1}_j=max(0,\gamma^k_j-c_k f^k({\bf r_j})).
\end{equation}

The vector $\gamma^k$ is set at 
the beginning of the iterative procedure to zeros. 
The unconditional minimization of $L_{c_k}$ can be performed by using any known minimization method. 
We adapted gradient search method because of its simplicity. 

\begin{figure}[bthp]
\centering
\includegraphics[width=0.8\hsize]{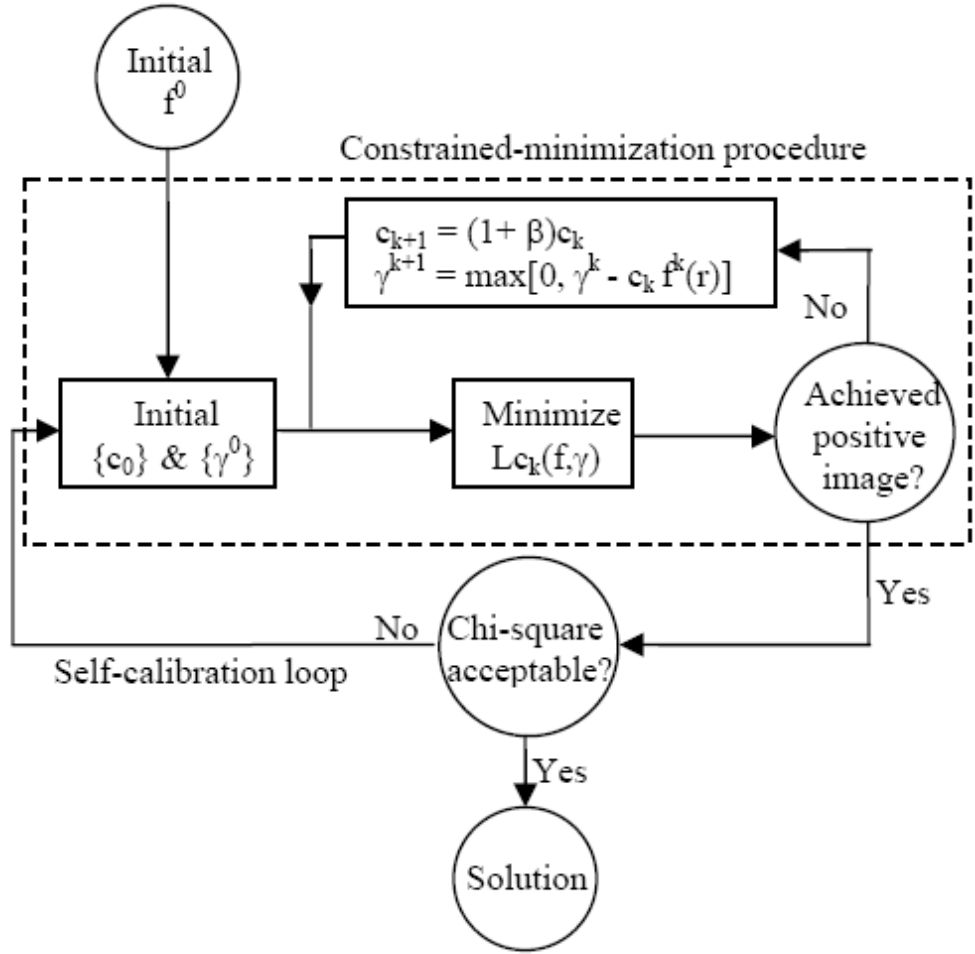}
\caption{A flow chart of the image reconstruction algorithm is shown
  in this figure. The dashed-line rectangle encloses the constrained-minimization procedure. The outer loop (outside this dashed-line rectangle) represents the self-calibration loop. As shown in this flow chart, the self-calibration loop uses the most recent values of $f(r_j)$ as the initial values for $f^k(r_j)$, while setting $\{\gamma^k\}$ and $\{c_k\}$ to initial values, namely, $\{\gamma^0\}$ and $\{c_0\}$ respectively. 
}
\label{flowChart}
\end{figure}

The conditional minimization problem (Equations~\ref{E1} \&~\ref{E2}) is incorporated into a self calibrating iterative loop where the output of the previous minimization is used as the initial approximation for the next minimization step. A Gaussian brightness distribution minimizing $\chi^2_{V^2}$ 
is assumed as the initial approximation for the first minimization
loop. The resolution of the final images is constrained by the apodization of the Fourier spectrum with a Gaussian function of $\sigma$ $\sim$ 0.7 B$_{max}$. A flow diagram of this minimization algorithm is shown in Figure~\ref{flowChart}.

\clearpage
\clearpage
 
\subsubsection{Reconstructed images}
\begin{figure}[bthp]
\centering
\includegraphics[width=0.98\hsize]{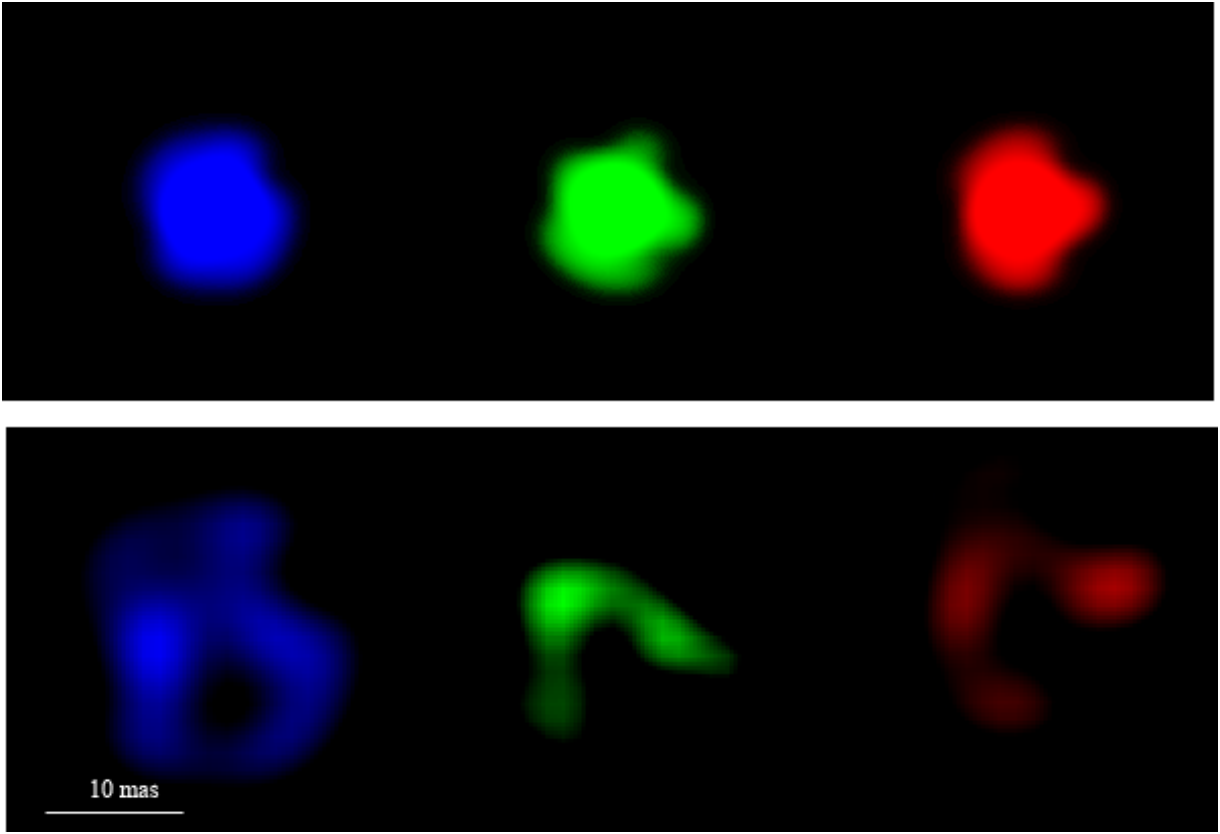}
\caption{Reconstructed near-infrared images of R Aqr are shown here.
  The blue, green and red color images (from left to right) represent
  1.51, 1.64 and 1.78 $\mu$m respectively. The top row shows
  reconstructed images and the bottom row shows primarily contributions from the shell since we subtracted a Gaussian function from the reconstructed images to remove the stellar component. 
North is up, east to the left.
}
\label{images}
\end{figure}

\begin{figure}[bthp]
\centering
\includegraphics[width=0.6\hsize]{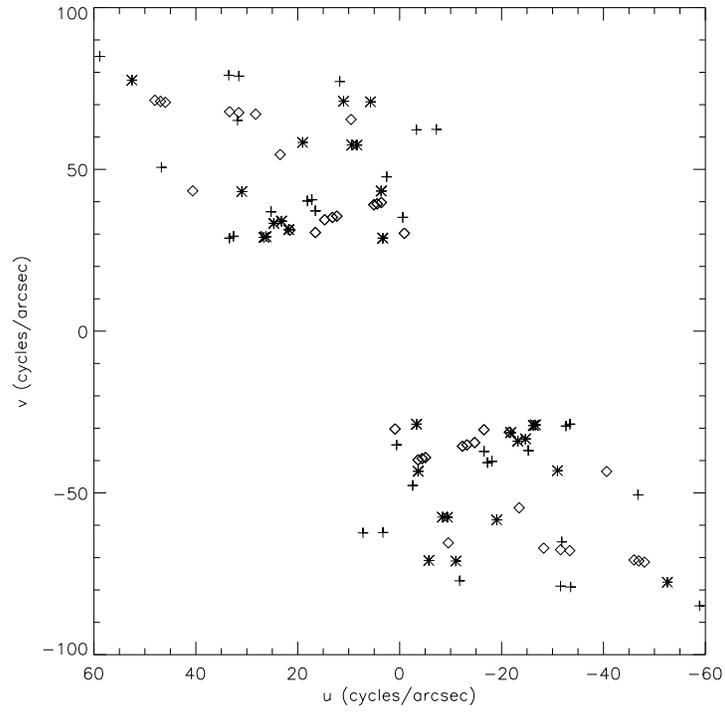}
\caption{The uv coverage of our observations are shown here. The symbols "plus", "star" and "diamonds" correspond to the observational wavelengths of 1.51$\mu$m, 1.64$\mu$m and 1.78$\mu$m respectively.  
}
\label{uv}
\end{figure}

The reconstructed images are shown in Figure~\ref{images}. The reduced
chi-squares for these images is estimated by summing the reduced
chi-squares for all visibility and the closure phase data points with values of $\chi^2_R$ = 1.96, 0.85 and 1.02 respectively at 1.51$\mu$m, 1.64$\mu$m and 1.78$\mu$m.
For clarity, we subtracted the central star contribution from these
reconstructed images and these are also shown in this figure. These
latter images clearly show an asymmetric shell with at least three
clumpy features. 
The uv coverage of our observations in shown in Figure~\ref{uv}.

\clearpage

\clearpage
\clearpage

\begin{figure}[bthp]
\centering
\includegraphics[width=0.98\hsize]{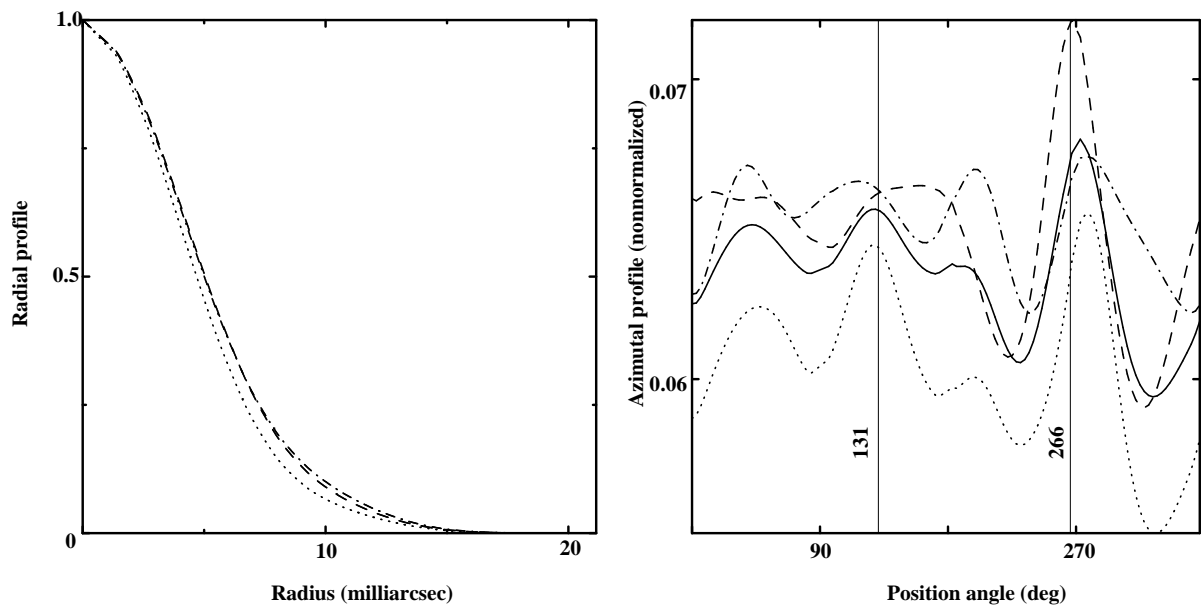}
\caption{{\bf Left:} The radial profiles of the azimuthally averaged,
  reconstructed images are plotted here. The dot-dashed, dotted and
  dashed lines correspond to 1.51 $\mu$m, 1.64 $\mu$m and 1.78 $\mu$m
  respectively. 
  {\bf Right:} The azimuthal profiles of the radially averaged, reconstructed images are plotted here. The dot-dashed, dotted and
  dashed lines correspond to 1.51 $\mu$m, 1.64 $\mu$m and 1.78 $\mu$m
  respectively. The continuous curve corresponds to the mean values.
 The azimuthal profiles clearly show at least three peaks. Two of those features correspond to off-axis point source solutions of the three component model presented in section 4.5.
}
\label{radial}
\end{figure}

Two of those clumpy features are roughly seen at positions where we
obtained an off-axis model feature for our three component models (section 4.5). Figure~\ref{radial} shows the radial and azimuthal profiles of these reconstructed images. 
The relative surface brightness of the shell with respect to the peak brightness of the image at 1.51, 1.64 and 1.78 $\mu$m is 7.8\%, 4.4\% and 6.4\% respectively. The shell is brighter at 1.51 $\mu$m and 1.78 $\mu$m than at 1.64 $\mu$m. 
This signifies the presence of water molecules in the shell. However, non-zero shell brightness at 1.64 $\mu$m signifies contributions from scattering as well as dust emissions at these three wavelengths. 

\clearpage
\clearpage

\subsubsection{Simple model}
\begin{figure}[bthp]
\centering
\includegraphics[width=0.4\hsize]{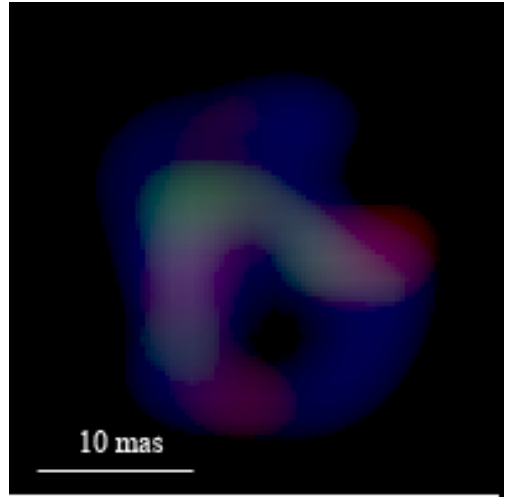}
\caption{The three wavelength images are composed and shown in this figure. The blue, green and red colors represent 1.51, 1.64 and 1.78 $\mu$m respectively. North is up, east to the left. 
}
\label{images2}
\end{figure}

The three wavelength images are composed and shown in Figure~\ref{images2}. The blue, green and red colors represent 1.51, 1.64 and 1.78 $\mu$m respectively. The 1.51$\mu$m image shows a bipolar feature. Moreover, the features at 1.64 and 1.78$\mu$m are somewhat consistent with this feature seen at 1.51$\mu$m. 
Interestingly the axis of symmetry of this bipolar feature is roughly parallel to the large scale astrophysical jets of R Aqr as well as the jets seen in SiO maser observations (Figure~\ref{R_AqrH2O}). We propose a simple speculative model in which the astrophysical jets to the north-east and south-west compress the material in these directions giving rise to the feature in Figure~\ref{images2}.
Additional measurements are essential to validate this simple model. If this hypothesis is true, then these observations are key steps in the understanding of the formation of bipolar nebulae.

\clearpage
\clearpage

\section{SiO Masers}

   The molecular envelopes of oxygen rich  Miras frequently exhibit SiO masers \citep{Reid97} in the region just interior to where the molecular gas condenses into dust.  The masers are strongest at the locations where the line of sight is tangent to the region giving rise to the masers, as these are the locations where the long paths through constant radial velocity gas needed to develop sufficient gain in the masers occur.  The masers appear as partial rings of bright spots centered on the star.


The IOTA observations (table 1) are taken in the early declining phase of one pulsation cycle, while the VLBA data were taken in the mid-rise of the next.  However, the travel times are such that a shock emerging at some time, t=0, through the photosphere will arrive at the SiO/molecular shell about one pulsation period later. 
Furthermore, monitoring of several maser shells reported in \citet{Diamond03}, \citet{Cotton04}, and \citet{Cotton06} show a complex interaction of the maser distribution with the pulsation cycle with variations of the maser distribution only weakly coupled to individual pulsation cycles.  Thus, the distribution of masers appears to be accumulated over several pulsation cycles; we assume that the separation of the IR and SiO observations by a cycle does not affect our interpretation of the results. 

    Interferometric observations of R Aqr in the $\nu$=1 and $\nu$=2, J=1-0 rotational transitions of SiO at 43.1 and 42.8 GHz are reported in \citet{Boboltz97}, \citet{Hollis00}, \citet{Hollis01}, \citet{Cotton04} and \citet{Cotton06}.
The average SiO ring diameter reported by \citet{Cotton06} for R Aqr
is 31.5 mas or somewhat larger than the H$_2$O shell diameter of 25.2 mas given in Table 2 or shown in Figure \ref{images}. The ratio of the size of the SiO shell to water shell is 1.25. Thus, the masers appear at the outer edge of the molecular envelope as reported in \citet{Perrin04} and \citet{Cotton06} for Mira and U Ori. 

\begin{figure}
\centerline{
\includegraphics[height=3.5in]{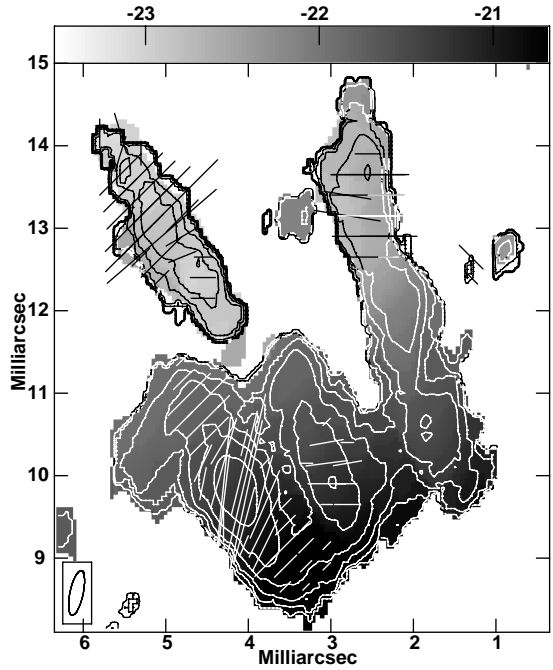}
}
\caption{
Contours of integrated Stokes' I at powers of 2 times 0.1 Jy/beam of
the SiO J=1-0, $\nu=$2 maser in R Aqr.
Superposed are vectors with length proportional to the integrated
linear polarization and orientation of the E-vectors.
The gray-scale gives the flux weighted average velocity whose scale 
is given by the wedge at the top of the plot.
The resolution is shown in the lower left corner.
}
\label{R_Aqr1PCntrVelA}
\end{figure}

\begin{figure}
\centerline{
\includegraphics[height=3.5in]{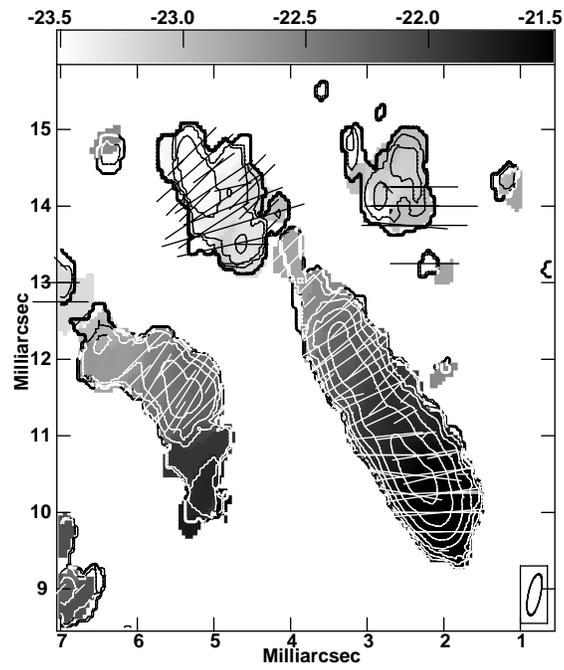}
}
\caption{
Like Figure \ref{R_Aqr1PCntrVelA} except for the SiO J=1-0,$\nu=$1 transition.
The resolution is shown in the lower right corner.
}
\label{R_Aqr2PCntrVelA}
\end{figure}

The SiO masers shown in Figure \ref{geometry} appear well outside of the $H_2O$ shell.  This suggests that the molecular region of the circumstellar envelope is more extended than given by the simple model fit or Figure ~\ref{images}. Figures \ref{R_Aqr1PCntrVelA} and \ref{R_Aqr2PCntrVelA} show velocity gradients in the outer regions of the molecular envelope.  Since these measured velocities are primarily tangential to the molecular envelope, they do not necessarily represent an overall expansion of the envelope but may well be part of a dust expulsion event as the gas giving rise to the masers condenses into silicate dust.

   The observations of \citet{Hollis01} and \citet{Cotton04} show that around 2001, and before, there appeared to be either general differential rotation of the envelope or a rotating molecular ring.
This feature, if interpreted as rotation, would have a period of ~22
sin(i) years where i is the inclination of the orbit.  The projected position angle of the ring on the sky is 80/260 degrees with the west side approaching \citep{Cotton04}.  \citet{Hollis00} interpret this systematic motion as being the result of tidal effects on the
molecular envelope during the periastron passage of the companion past
the Mira. This systematic motion was reported in observations between 1996 and 2001 but subsequent observations in September 2004 \citep{Cotton06} show no evidence for systematic rotation.

Figure \ref{R_AqrH2O} shows substantial differences from earlier observations given in \citet{Cotton06}. 
SiO masers around AGB stars generally consist of localized features forming a partial ring.
Turbulence will reduce lengths of the line of sight through constant
radial velocity gas, reducing the gain and hence the brightness of the
maser spots.
A ring--like structure is hard to determine from Figure
\ref{R_AqrH2O}, suggesting that the molecular envelope is more
turbulent than usual. 
In Miras better studied as a function of pulsation phase, no clear
relationship has been found between pulsation phase and SiO maser
structures \citep{Diamond03, Cotton06} and clear, if partial, rings are the norm.
In particular, both transitions show prominent jet--like features
to the north-east containing velocity gradients at position angles between 20 and 35
degrees. 
The velocity of these features becomes increasingly far from systemic
with increasing distance from the star.
Close-up views of these regions with polarization vectors are shown in
Figures \ref{R_Aqr1PCntrVelA} and \ref{R_Aqr2PCntrVelA}.
The orientations of the features seen in these figures are quite
different from the position angles (80, -100 degrees) of the 2001
features. 
The polarization vectors seen in these plots are mostly perpendicular
to the direction of the feature indicating that the magnetic field  is 
predominantly along the feature (see discussion in \citet{Cotton06}. 
Thus, the range of orientations of the jet--like features means that
it is unlikely that the magnetic fields in these features are
part of a large scale, persistent dipolar field.

The jet--like features seen in 2006 are approximately parallel to the
large scale jet from the companion.
In principle, the features could have been formed by the large
scale jet. 
However, the 2001 jet--like features show little relationship to
either the direction of the large scale jet or to the derived plane of
the orbit of the companion shown in Figure \ref{geometry}.
Furthermore, the orbit derived for the well established companion
places it far from the AGB star just prior to 2001.
Thus, any connection between the companion star or its jet and the SiO
features is doubtful.

\clearpage

\clearpage

\section{Conclusions}

In order to characterize the observed asymmetry reported in our
earlier work \citep{Ragland06}, we carried out follow-up observations
of selected Mira stars at near-infrared wavelengths. In this article,
we presented observations of the nearest known symbiotic Mira, R Aqr,
from the IOTA imaging interferometer and from the VLBA.

The reconstructed near-infrared images of R Aqr are reported for the
first time at three wavelengths namely, 1.51, 1.64 and 1.78 $\mu$m. The near-infrared images suggest a clumpy molecular shell around the central Mira star. The observed data could also be modeled by a three component model consisting of a symmetric central star surrounded by a water shell with a radius of about 2.25R$_*$, and an off-axis compact feature at about 2R$_*$ at a position angle of 131$^{\rm o}$ (an alternate model exists wherein the off-axis feature is at about R$_*$ at a position angle of -94$^{\rm o}$). 
However, the companion feature solution is inconsistent with the orbital parameters reported in the literature \citep{Hollis97}. We conclude that our observations are best explained with a clumpy, extended $H_2O$ circumstellar envelope.

The ratio of the size of water shell to the size of the central star is 2.25. The ratio of the size of SiO shell to that of water shell is 1.25. Thus, the masers appear at the outer edge of the molecular envelope as reported for other Mira stars in the literature. 

The SiO maser emission around R Aqr shows large scale turbulence in
the molecular envelope with prominent jet--like features towards the
north--east. 
These features are roughly parallel to the large scale jet from the more distant companion star discussed by \citet{Hollis97}. 
However, similar features seen in 2001 \citep{Cotton04,Hollis01} have a very different orientation from either the jet or
the inferred orbit of the companion so any connection between the SiO
features and the companion or its jet is doubtful.
The magnetic fields appear to be oriented parallel to the features
but due to the range of orientations seen in these features, their
magnetic fields are not part of a persistent, large scale dipole field. 

Further investigation and observations will be required to understand the details of this complex
and interesting system. 

\acknowledgments
This work was primarily supported by the NSF through research grant AST-0456047.
The IOTA is principally supported by the Smithsonian 
Astrophysical Observatory and the Univ. of Massachusetts. The National 
Radio Astronomy Observatory (NRAO) is operated by Associated 
Universities Inc., under cooperative agreement with the NSF. 
We thank Julien Montillaud and Julien Dejonghe for their support role in obtaining IOTA observations.
This research has made use of NASA's Astrophysics Data System Bibliographic
Services and light curve databases from AAVSO and AFOEV.

\end{document}